\documentclass[10pt]{article}
\usepackage{amsmath}
\begin{document}
%
\title{\bf Statistical Theory of Spin Relaxation and Diffusion in Solids 
\thanks{Journal of Low Temperature Physics 
(2006)(to be published)}} 
\author{
A. L. Kuzemsky \thanks {E-mail:kuzemsky@theor.jinr.ru;
http://theor.jinr.ru/~kuzemsky}
\\
{\it Bogoliubov Laboratory of Theoretical Physics,} \\
{\it  Joint Institute for Nuclear Research,}\\
{\it 141980 Dubna, Moscow Region, Russia.}}
\date{}
\maketitle
%
%
%
%
%
%
%
%
%
%
%
%
%
%
%
%
%
%
%
%
%
%
%
%
%
%
%
%
%
%
%
\begin{abstract}
A comprehensive theoretical description is given for the spin relaxation and diffusion in solids.
The formulation is made in a general statistical-mechanical way.
The method of the nonequilibrium statistical operator (NSO) developed by D. N. Zubarev is
employed to analyze a relaxation dynamics of  a spin  subsystem.  
Perturbation of this subsystem in solids may produce a nonequilibrium state which  is  then  relaxed
to an equilibrium state due to the interaction between the particles or with a thermal bath (lattice).
The generalized kinetic equations  were derived previously for a system weakly coupled to a thermal bath
to elucidate the nature of transport and relaxation processes.  In this paper, these results are used
to describe the relaxation and diffusion of nuclear spins in solids. The aim is to formulate 
a successive and coherent microscopic description of the nuclear magnetic relaxation and diffusion in solids.
The nuclear spin-lattice relaxation is considered and the Gorter relation is derived. As an example,
a theory of spin diffusion of the nuclear magnetic moment in dilute alloys ( like Cu-Mn) is developed. It is
shown  that due to the dipolar interaction between host nuclear spins and impurity spins, a nonuniform
distribution in the host nuclear spin system will occur and consequently the macroscopic relaxation time
will be strongly determined by the spin diffusion. The explicit expressions for the relaxation time in certain
physically relevant cases are given.

\end{abstract}
%
%
\section{INTRODUCTION}
%
%
%
%
For many years there has been considerable interest, experimental  and theoretical, in relaxation processes occurring 
in various spin systems, especially the nuclear spin systems in solids and 
liquids~\cite{cas37,cas39,bloc,gor,bl49,kt,red56,red57,red59,hsl,ca60,sher,sherp,blo,kla,ko,ko64,sten,cas64,hanh,a63,ak,bu65,sa66,red66,lo67,rob,jen68,mora,hm,hm1,gold,abr,wolf,sl,lou,ab,bo04}.
In ordinary spin resonance experiments, spins are subject to an applied magnetic field $h_{0}$ and make
a precessional motion around it. Local fields produced by interactions of the spins with their environments
act as relatively weak perturbations to the unperturbed precessional motion. In quantum-mechanical language,
the external field gives rise to the Zeeman levels for each spin and the interactions are perturbations to these quantum
states. In a nuclear-magnetic resonance (NMR) experiment, the nuclear spin system absorbs energy from the externally
applied radio-frequency field and transfers it to the thermal bath or reservoir provided by the lattice through
the spin lattice interaction. The coupled nuclear spins in a solid with very slow spin-lattice relaxation time $T_{1}$
comprise a  quasi-isolated system which
for many purposes can be treated by thermodynamic methods. 
The spin-spin
relaxation time is denoted by $T_{2}$. The other system, called the lattice, contains all other
degrees of freedom, phonons, translational motion of conduction electrons, etc. It is at a temperature   $T$ that it
is considered stable. 
A macroscopic approach to the description of
magnetic relaxation was proposed by Bloch~\cite{bloc}. He proposed a phenomenological equation describing the motion of
nuclear-spin system subjected to both a static and a time varying magnetic field 
$$ \frac{d\vec{M}}{dt} = \gamma \vec{M}\times \vec{h} - \frac{M_{x}}{T_{2}}\vec{i} - \frac{M_{y}}{T_{2}}\vec{j} +
\frac{M_{0} - M_{z}}{T_{1}}\vec{k}$$
where the external field $\vec{h}$ is taken to be of the form $\vec{h} = h_{0}\vec{k} + 2h_{1}(t) cos \omega t \vec{i}$.
This equation successfully describes a wide variety of 
magnetic resonance experiments, although to obtain a valid description of low-frequency phenomena, it is necessary to modify the original
equation so that relaxation takes place toward the instantaneous magnetic field.
In an NMR experiment, the absorption of energy from the applied rf field produces either an increase in the energy of 
the spin system  or a transfer of energy from the spin system to the lattice. The latter process requires a time 
interval of the order of spin-lattice relaxation time $T_{1}$. The characteristic time $T_{2}$ determines the relaxation
of the transversal spin components due to the spin-spin interactions.\\
The relaxation processes in spin systems have been
investigated by a number of authors~\cite{red57,blo,hanh,a63,ak,bu65,red66,lo67,rob,jen68,abr,ab} to obtain qualitative 
and quantitative information about irreversible spin-spin and spin-lattice processes in spin systems. The method of many of these 
papers was to develop an equation of motion for the reduced density matrix~\cite{sher,sherp,ko,ko64} describing the spin system, and was found to be
most useful when the perturbation responsible for the relaxation of the spin system had a very short correlation time. In the equation-of-motion
approach, the specification of the initial conditions involves the assumption of some explicit form for the density matrix describing the system
(the system includes both the spin and its surroundings, which in the case studied below will be the conduction electrons in a metal).
This problem is very
attractive  from the point of view of irreversible statistical mechanics since a general model of magnetic resonance
consists of a driven system of interest in interaction with a heat bath. Stenholm and ter Haar~\cite{sten} have analyzed the basic assumptions
which are necessary for statistical-mechanical derivation of the Bloch equation and the role of the thermal bath.\\
An important concept in the interpretation of spin-lattice
relaxation phenomena was provided by the thermodynamic theory of Casimir and Du Pre~\cite{cas37}. They 
considered the magnetic crystal to be composed of two subsystems which could be assigned two different temperatures.
One subsystem contained the magnetic degrees of freedom. The other subsystem, called the lattice, contains all
other degrees of freedom. Then, the idea of spin temperature
was extended and several distinct temperatures for magnetic subsystems (Zeeman, dipole-dipole, etc. ) 
were introduced~\cite{gold}. (Note, however, some special exclusions~\cite{pin}).
In general, the state of the total system to be composed of a few subsystems may be described approximately by
a density matrix of the form
\begin{equation}\label{eq0}
 \rho \sim \exp [ - (H_{1}/kT_{1}) -  (H_{2}/kT_{2}) - (H_{3}/kT_{3}) \ldots ], \nonumber
\end{equation}
with a number of quasi-invariant energies $Tr(H_{i})$ and a number of distribution parameters $T^{-1}_{i}$.
Nuclear relaxation in weak applied fields was first treated by Redfield~\cite{red59} 
and by Hebel and Slichter~\cite{hsl}, using the idea of spin temperature. Redfield theory is the semiclassical 
density operator theory of spin relaxation.\\
It was Bloembergen~\cite{bl49} who first  formulated that the magnetization of spins in a rigid lattice could 
be spatially transported by means of the mutual flipping of neighboring spins due to dipole-dipole interaction.
This idea permitted one to explain the significant influence of a small concentration of paramagnetic impurities
on spin-lattice relaxation in ionic crystals.
He used a quantum-mechanical treatment ( first-order perturbation theory) and showed that the transport equation
for magnetization was a diffusion equation. In this simple approximation he calculated the diffusion constant $D$.
In other words, we can roughly represent the relaxation dynamics  as
\begin{eqnarray}\label{eq104}
\frac{\partial \langle I^{z} (\vec{r})\rangle  }{\partial t} = - A (\vec{r})[\langle I^{z} (\vec{r})\rangle - 
\langle I^{z} (\vec{r})\rangle^{0} ] + D(\vec{r})\nabla^{2} \langle I^{z} (\vec{r})\rangle \\ \nonumber
\frac{\partial \langle I^{z} (\vec{r})\rangle  }{\partial t} = - \frac{\langle I^{z} (\vec{r})\rangle - 
\langle I^{z} (\vec{r})\rangle^{0}}{T_{1}} \\ \nonumber
\frac{1}{T_{1}} \propto \frac{1}{T^{SL}_{1}} + \frac{1}{T^{D}_{1}} \nonumber
\end{eqnarray}
where $I^{z}$ is the z-component of the nuclear spin operator.\\
Since then, many authors have formulated the general theory of the spin relaxation processes in solids from the standpoint
of statistical mechanics or irreversible thermodynamics. An improvement in the general formulation of the theory was achieved by 
Kubo and Tomita~\cite{kt} in their treatment of magnetic resonance absorption via a linear  theory of irreversible processes. In this theory the important 
quantities are frequency-dependent susceptibilities which are expressed in terms of spin correlation functions. Buishvili~\cite{buis65} developed a 
quantum-statistical theory of the dynamic polarization of nuclei by taking into account diffusion of nuclear spins as well as dipole 
interaction of electron spins. 
Buishvili and Zubarev~\cite{bu65} developed a successive theory
of spin diffusion in  crystals. The nuclear diffusion in diamagnetic solids with paramagnetic impurities
was analyzed by the method of the statistical operator for nonequilibrium systems. The Bloembergen equation~\cite{bl49}, whose 
coefficients are explicitly expressed through certain correlation functions, was obtained. 
The theory of nuclear spin diffusion in the ferromagnets of certain type was considered in Ref.~\cite{bui65}. The theory of the dynamic polarization of 
nuclei and nuclear relaxation for the case of strong saturation was analyzed in Ref.~\cite{bbz}. The influence of a strong NMR saturation on
spin diffusion was considered in Ref.~\cite{buzv}. The time of spin-lattice relaxation was calculated for nuclei when spin diffusion was taken
into account under conditions of a strong NMR saturation. The influence of exchange interactions between nuclear spins on the dynamic 
polarization of nuclei was considered in Ref.~\cite{buzv68}.
Borckmans and Walgraef~\cite{bowa} formulated a theory
of Zeeman and dipolar energy diffusion in paramagnetic spin systems in the frame the general theory of irreversible processes developed by 
Prigogine and co-workers. Buishvili and Giorgadze~\cite{bug69} investigated general theory of spin diffusion within the nonequilibrium
statistical operator approach. A consistent quantum statistical investigation of saturation of a nonuniformly broadened EPR line was carried out 
in Ref.~\cite{bzkh69} by taking into account spectral diffusion and the dipole-dipole reservoir.
Role of the flip-flip and flip-flop transitions for the dynamic 
polarization of nuclei was analyzed in Ref.~\cite{bug70}. The theory of spin-lattice relaxation in crystals with paramagnetic impurities was discussed
in paper of Bendiashvili, Buishvili and Zviadadze~\cite{bbz70}. The analysis of the role of the interaction between a few subsystems for the
construction of the nonequilibrium density matrix was discussed from a general point of view by Buishvili and Zviadadze~\cite{bz70}. 
The application of the nonequilibrium statistical operator method to the case of relaxation in dilute alloys has been considered 
by Fazleev~\cite{faz,faz1}.
The influence of relation between 
thermal capacities of nuclear spin subsystem and  the reservoir of electron spin-spin interaction on spin kinetics, especially in the low-temperature case when the spin 
polarization of subsystems are high enough was analyzed in detail by Tayurskii~\cite{tay} for the case of insulators.\\
Robertson~\cite{rob} derived an equation of motion for the total
magnetic moment of a system containing a single species of nuclear spins in an arbitrarily time-dependent external 
magnetic field. He derived a generalization of Bloch's phenomenological equation for a magnetic resonance.
In papers~\cite{a63,ak}, the general quantum-statistical-mechanical approach to the problem of spin resonance and relaxation
which utilized a projection operator technique was developed. From the Liouville equation for the combined system of the spin
subsystem and the thermal bath a non-Markoffian equation for the time development of the statistical density operator
for the spin system alone was derived. The memory effects were taken into account in the application of the method of the statistical operator 
for nonequilibrium systems to magnetic relaxation problem by Nigmatullin and Tayurskii~\cite{nigm}.
Romero-Ronchin, Orsky and Oppenheim~\cite{rom}    used a projection operator technique for derivation of the 
Redfield equations~\cite{red57}.  In their paper~\cite{rom},
the relaxation properties of a spin system weakly coupled to lattice
degrees of freedom were described using an equation of motion for the
spin density matrix. This equation was derived using a general weak
coupling theory which was previously developed. To second order in
the weak coupling parameter, the results are in agreement with those
obtained by Bloch, Wangsness and Redfield, but the derivation does not
make use of second order perturbation theory for short times. The authors claim that the
derivation can be extended beyond second order and ensures that the spin
density matrix relaxes to its exact equilibrium form to the appropriate
order in the weak coupling parameter.\\
In this work, we present  a complementary theory which examines the relaxation dynamics of a spin system  in the approach of the 
nonequilibrium  statistical operator. It uses a general formalism from a previous study
 for a system  that is in contact  with a thermal bath
( a "lattice" ) and relax to the equilibrium state. 
The aim of this paper is to show how the general theory of irreversible processes allows a theoretical study of
such phenomena without postulated equations of phenomenological assumptions.
One of our purposes in this paper is to present a unified statistical mechanical treatment of spin relaxation 
and spin diffusion phenomena. The transport of nuclear spin energy in a lattice of paramagnetic spins with
magnetic dipolar interaction  plays an important role in many relaxation processes. In this paper
the microscopic derivation of an expression for the longitudinal relaxation time of bulk
metal nuclear spins by dilute local moments  is performed taking into account spin diffusion processes within the
nonequilibrium statistical operator approach. \\ In the next section, we establish the notation and briefly  present  the main ideas of the NSO approach. 
This section includes a
short summary of the derivation of the generalized kinetic and rate equations with the NSO method. Section 2 serves
as an extended introduction to the present paper.
In section 3, the dynamics of the nuclear spin system is analyzed.  We 
consider the application of the established equations to the derivation of the relaxation equations for
spin systems. Special attention is given to the problem of spin relaxation and diffusion in section 4. 
The case of nuclear spin diffusion in dilute magnetic alloys is discussed in some detail in section 4.2. 
The final section contains some concluding remarks concerning the results obtained.
\section{BASIC NOTIONS}\label{bn}
The statistical mechanics of irreversible processes in solids, liquids, and complex
materials like a soft matter are at the present time of much interest~\cite{grand87,grand88,kubo92,kubo91}. 
The central problem of nonequilibrium statistical mechanics is to derive a set of equations which
describe irreversible processes from the reversible equations of motion~\cite{grand88,kubo91}.
The consistent calculation of transport coefficients
is of particular interest because  one can get information on the microscopic
structure of the condensed matter.
During the last decades, a number of schemes have been 
concerned with a more general and consistent approach to transport 
theory~\cite{grand88,zub74,macl89,rz}.
These approaches, each in its own way, lead us to substantial advances in the understanding of the
nonequilibrium behavior of many-particle classical and quantum systems. This field is very active and
there are many aspects to the problem~\cite{leb}. 
Our purpose here is to discuss the derivation, within the formalism of the nonequilibrium
statistical operator~\cite{zub74}, of the generalized transport and kinetic equations. On this basis
we have derived, by statistical mechanics methods, the kinetic equations for a system 
weakly coupled to a thermal bath~\cite{kuz05}. Our motivation for presenting this alternative derivation is based on the
conviction that the NSO method provides some advantages in displaying the physics of the relaxation processes. 
%
%
%
\subsection{Outline of the Nonequilibrium Statistical Operator Method}\label{nso}
%
%
In this section, we  briefly   recapitulate 
the main ideas of the  nonequilibrium statistical operator approach~\cite{zub74,kuz05} for the sake of a self-contained 
formulation. 
The precise definition of the nonequilibrium state is quite difficult
and complicated, and is not uniquely specified. Since it is virtually impossible and impractical to
try to describe in detail  the state of a complex macroscopic system in the nonequilibrium state, the
method of reducing the number of relevant variables was widely used. A large and important class
of transport processes can  reasonably be  modelled in terms of a reduced number of macroscopic
relevant variables. There are different time scales and different sets of the relevant variables~\cite{bog,bog1},
e.g. hydrodynamic, kinetic, etc. 
The most satisfactory and workable approach 
to the construction of  Gibbs-type ensembles for the nonequilibrium systems, as it appears to the writer, is the method of 
nonequilibrium statistical operator (NSO) developed by D. N. Zubarev~\cite{zub74}. The NSO method
permits one to generalize the Gibbs ensemble method   to the nonequilibrium case and 
construct a nonequilibrium statistical operator which  enables one to obtain the transport equations and
calculate the kinetic coefficients in terms of correlation functions, and which, in the case of
equilibrium, goes over to the Gibbs distribution. Although this method is well known, we shall
briefly recall it, mostly in order to introduce the notation needed in the following.\\
The NSO method sets out  as follows. The irreversible processes which can be considered as a reaction
of a system on mechanical perturbations can be analyzed by means of the method of linear reaction
on the external perturbation~\cite{kubo91}. However, there is also a class of irreversible
processes induced by  thermal perturbations due to the internal inhomogeneity of a system. Among them
we have, e.g., diffusion, thermal conductivity, and viscosity. In   certain approximate schemes it is
possible to express such processes by  mechanical perturbations which artificially induce similar nonequilibrium
processes. However, the fact is that the division of perturbations into mechanical and thermal ones is
reasonable in the linear approximation only. In the higher approximations in the perturbation, 
mechanical perturbations can  effectively lead to the appearance of  thermal perturbations.\\ The
NSO method permits one to formulate a workable scheme for description of the statistical mechanics of
irreversible processes which include the thermal perturbation in a unified and coherent fashion.
To perform this, it is necessary to construct  statistical ensembles representing the macroscopic conditions
determining the system. Such a formulation is quite reasonable if we consider our system for a suitable large time.
For these large times the particular properties of the initial state of the system are irrelevant and
the relevant number of variables necessary for  description of the system reduces substantially~\cite{bog}.\\
The basic hypothesis is that after small
time-interval $\tau$ the nonequilibrium distribution is established. Moreover, it is supposed that
it is weakly time-dependent by means of its parameter only. Then the statistical operator $\rho$ for
$t \geq \tau$ can be considered as an "integral of motion" of the quantum Liouville equation
\begin{equation}\label{eq1}
 \frac{\partial \rho }{\partial t} - \frac{1}{i \hbar}[\rho, H ] = 0
\end{equation}
Here $\frac{\partial \rho }{\partial t}$ denotes time differentiation with respect to the time variable
on which the relevant parameters $F_{m}$ depend. It is important to note once again that $\rho$ depends
on $t$ by means of $F_{m}(t)$ only. We may consider that the system is in thermal, material, and mechanical contact
with a combination of thermal baths and reservoirs  maintaining the given distribution of  parameters
$F_{m}$.  For example,  it can be the densities of energy, momentum, and particle  number for the
system which is macroscopically defined by given fields of temperature, chemical potential and
velocity. It is assumed that the chosen set of parameters is sufficient to characterize  macroscopically
the state of the system. The set of the relevant parameters are dictated by the external
conditions for the system under consideration and, therefore, the term $\frac{\partial \rho }{\partial t}$
appears as the result of the external influence upon the system. Due to this influence precisely, the
behavior of the system is nonstationary.\\ In order to describe the nonequilibrium process, it is
necessary also to choose the reduced set of relevant operators $P_{m}$, where $m$ is the index ( continuous 
or discrete). In the quantum case, all operators are considered to be in the Heisenberg representation
\begin{equation}\label{eq2}
 P_{m}(t) = \exp ( \frac{iHt}{\hbar})P_{m}\exp ( \frac{-iHt}{\hbar})
\end{equation}
where $H$ does not depend on the time. The relevant operators may be scalars or vectors. The equations
of motions for  $P_{m}$ will lead to the suitable "evolution equations"~\cite{zub74}. In the quantum
case
\begin{equation}\label{eq3}
\frac{\partial P_{m}(t)  }{\partial t} - \frac{1}{i \hbar}[P_{m}(t) , H ] = 0. 
\end{equation}
The time argument of the operator $P_{m}(t)$ denotes the Heisenberg
representation with the Hamiltonian $H$ independent of time.
Then we suppose that the state of the ensemble
is  described by a nonequilibrium statistical operator which is a functional of $P_{m}(t)$
\begin{equation}\label{eq4}
  \rho(t ) = \rho \{\ldots P_{m}(t)  \ldots \}
\end{equation}
Then $\rho(t )$ satisfies the Liouville equation (\ref{eq1}). Hence the  quasi-equilibrium ( "local-equilibrium")
Gibbs-type distribution will have the form
\begin{equation}\label{eq5}
 \rho_{q} = Q^{-1}_{q} \exp \left(  - \sum_{m}F_{m}(t)P_{m}\right)
\end{equation}
where the parameters $F_{m}(t)$ have the meaning of time-dependent thermodynamic parameters,
e.g., of temperature, chemical potential, and velocity ( for the hydrodynamic stage), or the
occupation numbers of one-particle states (for the kinetic stage). The statistical
functional $Q_{q}$ is defined by demanding that the operator $\rho_{q}$ be normalized and  equal
to
\begin{equation}\label{eq6}
 Q_{q} = Tr \exp \left(  - \sum_{m}F_{m}(t)P_{m}\right)
\end{equation}
This description is still very simplified. There are various effects which can make the picture
more complicated. The quasi-equilibrium distribution is not necessarily close to the stationary
stable state. There exists another, completely independent method for choosing a suitable
quasi-equilibrium distribution~\cite{grand87,grand88,macl89,zbig}. For the state with the extremal
value of the informational entropy~\cite{grand88,macl89}
\begin{equation}\label{eq7}
  S = - Tr ( \rho \ln \rho),
\end{equation}
provided that
\begin{equation}\label{eq8}
   Tr ( \rho P_{m} ) = \langle P_{m}\rangle_{q}; \quad  Tr   \rho  = 1,
\end{equation}
it is possible to construct a suitable quasi-equilibrium ensemble.
Then the corresponding quasi-equilibrium ( or local equilibrium ) distribution  has
the form
\begin{eqnarray}\label{eq9}
\rho_{q}   =   \exp \left( \Omega - \sum_{m}F_{m}(t)P_{m}\right) \equiv \exp ( - S(t,0))\\
\Omega = \ln Tr  \exp \left( - \sum_{m}F_{m}(t)P_{m}\right), \nonumber
\end{eqnarray}
where $S(t,0)$ can be called  the entropy operator.
The form of the quasi-equilibrium statistical operator was constructed  so as to
ensure  that the thermodynamic equalities for the relevant parameters $F_{m}(t)$
\begin{equation}\label{eq10}
  \frac{\delta \ln Q_{q}}{\delta F_{m}(t)} = \frac{\delta \Omega}{\delta F_{m}(t)} =  - \langle P_{m}\rangle_{q};
\quad \frac{\delta S}{\delta \langle P_{m}\rangle_{q} }  =  F_{m}(t) 
\end{equation}
are satisfied. It is clear that the variables $ F_{m}(t)$ and $\langle P_{m}\rangle_{q} $ are thermodynamically
conjugate.
Here the notation used is  $ \langle \ldots \rangle_{q} =  Tr ( \rho_{q}  \ldots )$.\\
By definition a special set of operators should be constructed which depends
on the time through the parameters $F_{m}(t)$ by taking the \emph{invariant part} of the operators
$F_{m}(t)P_{m}$ occurring in the logarithm of the quasi-equilibrium distribution, i.e.,
\begin{eqnarray}\label{eq11}
 B_{m}(t) = \overline{F_{m}(t)P_{m}} = \varepsilon \int^{0}_{-\infty} e^{\varepsilon t_{1}}
F_{m}(t+ t_{1})P_{m}(t_{1})dt_{1} = \\
F_{m}(t)P_{m} - \int^{0}_{-\infty} dt_{1} e^{\varepsilon t_{1}}\left(F_{m}(t+ t_{1})\dot{P}_{m}(t_{1}) +
\dot{ F}_{m}(t+ t_{1})P_{m}(t_{1}) \right)\nonumber
\end{eqnarray}
where $(\varepsilon \rightarrow 0)$ and 
$$ \dot{P}_{m} =  \frac{1}{i \hbar}[P_{m}, H ]; \quad  \dot{ F}_{m}(t) = \frac{d F_{m}(t)}{dt}.$$
The parameter $\varepsilon > 0$ will be set
equal to zero, but only after the thermodynamic limit has been taken. Thus, the invariant part is taken
with respect to the motion with Hamiltonian $H$.
The operation of taking the invariant part, of smoothing the oscillating terms, is used in the
formal theory of scattering~\cite{gell} to set the boundary conditions which exclude the advanced solutions
of the Schrodinger equation. 
The $\overline{P_{m}(t)}$ will be called the integrals ( or quasi-integrals ) of motion, although
they are conserved only in the limit $(\varepsilon \rightarrow 0)$. It is clear that for the Schrodinger equation
such a procedure excludes the advanced solutions by  choosing the initial conditions. In the
present context this procedure leads to the selection of the retarded solutions of the Liouville equation.\\ 
It should be noted that the same calculations can also be made with a  deeper
concept, the methods of quasi-averages~\cite{qbog,zub74}. 
\begin{equation}\label{eq14a}
 \frac{\partial \ln \rho_{\varepsilon}  }{\partial t} - \frac{1}{i \hbar}[ \ln \rho_{\varepsilon}, H ] = 
 - \varepsilon ( \ln \rho_{\varepsilon}  - \ln \rho_{q}),
\end{equation}
where $(\varepsilon \rightarrow 0)$ after the thermodynamic limit. 
The required nonequilibrium statistical operator is defined as
\begin{equation}\label{eq17}
\rho_{\varepsilon}  = \rho_{\varepsilon}(t,0)   =  \overline {\rho_{q}(t,0)} = 
\varepsilon \int^{0}_{-\infty} dt_{1} 
e^{\varepsilon t_{1}} \rho_{q}(t + t_{1}, t_{1})   
\end{equation}
Hence the nonequilibrium statistical operator can then be written in the form
%
\begin{eqnarray}
\label{eq18}
 \rho = Q^{-1} \exp \left(  - \sum_{m} B_{m} \right)   = \nonumber \\ \nonumber
\\ \nonumber Q^{-1} \exp \{  - \sum_{m} F_{m}(t) P_{m} + 
\sum_{m} \int^{0}_{-\infty}  dt_{1}e^{\varepsilon t_{1}} [ \dot{F}_{m}(t+ t_{1})P_{m}(t_{1}) + \nonumber \\
  F_{m}(t+ t_{1}) \dot{P}_{m}(t_{1})] \}  
\end{eqnarray}
Now we can rewrite the nonequilibrium statistical operator in the following useful form:
\begin{eqnarray}\label{eq18d}
    \rho (t,0) =  \exp \left(
 - \varepsilon \int^{0}_{-\infty} dt_{1} e^{\varepsilon t_{1}} \ln \rho_{q}(t + t_{1},  t_{1})  \right) = \nonumber \\
 \exp \overline{\left( \ln \rho_{q}(t,0)\right)} \equiv \exp \overline{\left( -S(t,0)\right)}
\end{eqnarray}
The average value of any dynamic variable $A$ is given by
\begin{equation}\label{eq18e}
  \langle A \rangle = \lim_{\varepsilon \rightarrow 0^{+}} Tr ( \rho (t,0) A )
\end{equation}
and is, in fact, the quasi-average. The normalization of the quasi-equilibrium distribution $\rho_{q}$ will
persists after taking the invariant part if the following conditions are required
\begin{equation}\label{eq18f}
  Tr ( \rho (t,0) P_{m} )= \langle P_{m}\rangle = \langle P_{m}\rangle_{q} ; \quad  Tr   \rho  = 1  
\end{equation}
Before closing this section, we shall mention some modification  of the "canonical"  NSO  method
which was proposed in Ref.~\cite{zbig} and which one has to take into account in a more accurate treatment
of transport processes.
%
%
\subsection{Transport and Kinetic Equations}\label{nsok}
%
It is well known that the kinetic equations are of great interest in the theory of transport
processes. Indeed, as it was shown in the preceding section, the main quantities involved are
the following thermodynamically conjugate values:
\begin{equation}\label{eq19a}
 \langle P_{m}\rangle = - \frac{\delta \Omega}{\delta F_{m}(t)};
\quad F_{m}(t)  = \frac{\delta S}{\delta \langle P_{m}\rangle }   
\end{equation}
The generalized transport equations which describe the time evolution of variables $\langle P_{m}\rangle$ and $F_{m}$
follow from the equation of motion for the $ P_{m}$, averaged with the nonequilibrium statistical
operator (\ref{eq18d}). It reads
\begin{equation}\label{eq19b}
 \langle \dot {P}_{m}\rangle = - \sum_{n} \frac{\delta^{2} \Omega}{\delta F_{m}(t)\delta F_{n}(t)}\dot{F}_{n}(t);
\quad \dot{F}_{m}(t)  =  \sum_{n} \frac{\delta^{2} S}{\delta \langle P_{m}\rangle \delta \langle P_{n}\rangle} \langle\dot {P}_{n}\rangle  
\end{equation}
The entropy production has the form
\begin{equation}\label{eq19c}
\dot {S}(t) = \langle \dot {S}(t,0)\rangle = - \sum_{m}\langle \dot {P}_{m}\rangle F_{m}(t) = -
\sum_{n,m}\frac{\delta^{2} \Omega }{\delta F_{m}(t)\delta F_{n}(t)}\dot{F}_{n}(t)F_{m}(t)
\end{equation}
These equations are the mutually conjugate and  with   Eq. (\ref{eq19a}) form a complete system of equations
for the calculation of values  $\langle P_{m}\rangle$ and $F_{m}$.
%
%
\subsection{ System in Thermal Bath: Generalized Kinetic Equations}\label{nsob}
%
In paper~\cite{kuz05} we derived  the generalized kinetic equations for  the system weakly coupled to a thermal bath.
Examples of such systems can be an atomic ( or molecular) system interacting with the electromagnetic
field it generates as with a thermal bath, a system of nuclear or electronic spins interacting with the lattice,
etc.
The aim was to describe  the relaxation processes in two weakly interacting subsystems, one of which is in the
nonequilibrium state and the other is considered as a thermal bath. The concept of thermal bath or heat reservoir,
i.e.,  a system that has effectively an infinite number of degrees of freedom, was not formulated precisely. A 
standard definition of the thermal bath is a heat reservoir defining a temperature of the system environment.
From a mathematical point of view~\cite{bog1}, a heat bath is something that gives a stochastic influence on the system
under consideration. In this sense, the generalized master equation~\cite{vh} is a tool for extracting the dynamics of
a subsystem of a larger system  by the use of a special projection techniques~\cite{zw}  or special 
expansion technique~\cite{fk}. The problem of
a small system weakly interacting with a heat reservoir has various aspects. 
Basic to the derivation of a
transport equation for a small system weakly interacting with a heat bath is a proper introduction of  model assumptions. 
We are interested here in the problem of
derivation of the kinetic equations for a certain set of  average values 
( occupation numbers, spins, etc.)  which characterize the nonequilibrium state of the system.\\
The Hamiltonian of the
total system is taken in the following form:
\begin{equation}\label{eq42}
 H = H_{1} + H_{2} + V,
\end{equation}
where
\begin{equation}\label{eq43}
 H_{1} = \sum_{\alpha} E_{\alpha}a^{\dagger}_{\alpha}a_{\alpha}; 
 \quad V = \sum_{\alpha,\beta}\Phi_{\alpha \beta}a^{\dagger}_{\alpha}a_{ \beta}, \quad \Phi_{\alpha \beta} = 
 \Phi^{\dagger}_{\alpha \beta}
\end{equation}
Here $H_{1}$ is the Hamiltonian of the small subsystem, and $a^{\dagger}_{\alpha}$ and $a_{\alpha}$ are the creation and annihilation second
quantized operators of quasiparticles in the small subsystem with energies $ E_{\alpha}$, $V$ is the
operator of the interaction between the small subsystem and the thermal bath, and $H_{2}$ is the
Hamiltonian of the thermal bath  which we do not write explicitly. The quantities $\Phi_{\alpha \beta}$ are
the operators acting on the thermal bath variables.\\ 
We assume that the state of this
system is determined completely by the set of averages
$\langle P_{\alpha \beta}\rangle = \langle a^{\dagger}_{\alpha}a_{ \beta}\rangle$ and the state of the thermal bath 
by $\langle H_{2}\rangle$,
 where $ \langle \ldots \rangle$ denotes the statistical average with the nonequilibrium statistical operator,
 which will be defined below.\\
We take the quasi-equilibrium statistical operator $\rho_{q}$ in
the form
\begin{eqnarray}\label{eq44}
\rho_{q} (t) =  \exp (- S(t,0)), \quad  S(t,0) = \Omega(t) + 
\sum_{\alpha \beta }P_{\alpha \beta }F_{\alpha\beta }(t) + \beta H_{2}\\ 
\Omega = \ln Tr \exp (- \sum_{\alpha \beta }P_{\alpha \beta }F_{\alpha\beta }(t) - \beta H_{2} ) \nonumber
\end{eqnarray}
Here $F_{\alpha\beta }(t)$ are the thermodynamic parameters conjugated with $P_{\alpha \beta }$, and $\beta$ is
the reciprocal temperature of the thermal bath. All the operators are considered in the Heisenberg
representation. The nonequilibrium statistical operator has the form 
\begin{eqnarray}\label{eq44a}
\rho (t) =  \exp (- \overline{S(t,0)}),\\
\overline{S(t,0)} = \varepsilon \int^{0}_{-\infty} dt_{1} e^{\varepsilon t_{1}} 
\left( \Omega(t + t_{1}) +  \sum_{\alpha \beta }P_{\alpha \beta }F_{\alpha\beta }(t) + \beta H_{2}  \right)
\nonumber
\end{eqnarray}
The parameters $F_{\alpha\beta }(t)$ are determined from the condition 
$\langle P_{\alpha \beta }\rangle = \langle P_{\alpha \beta }\rangle_{q}$.\\ In the derivation of the kinetic equations we  use
the perturbation theory in a "weakness of interaction" and  assume that the equality 
$ \langle \Phi_{\alpha \beta}\rangle_{q} = 0$ holds, while  other terms can be added to the renormalized energy of
the subsystem. 
For further considerations it is convenient to rewrite $\rho_{q}$ as
\begin{equation}\label{eq47}
 \rho_{q}  = \rho _{1}\rho_{2} = Q^{-1}_{q} \exp (- L_{0}(t)),
\end{equation}
where
\begin{eqnarray}\label{eq48}
\rho _{1} = Q^{-1}_{1}\exp \left(- \sum_{\alpha \beta } P_{\alpha \beta } F_{\alpha \beta }(t)\right); \quad
Q_{1} = Tr \exp \left( - \sum_{\alpha \beta } P_{\alpha \beta } F_{\alpha \beta }(t) \right)\\
\rho _{2} = Q^{-1}_{2} e^{- \beta H_{2} }; \quad Q_{2} = Tr \exp (- \beta H_{2})\\
Q_{q} = Q_{1}Q_{2}; \quad L_{0} = \sum_{\alpha \beta } P_{\alpha \beta } F_{\alpha \beta }(t) + \beta H_{2}
\end{eqnarray}
We now turn to the derivation of the kinetic equations. The starting point is the kinetic equations
in the following implicit form:
\begin{equation}\label{eq49}
 \frac{d \langle P_{\alpha \beta }\rangle }{d t} =  \frac{1}{i \hbar}\langle [  P_{\alpha \beta  } , H ]\rangle =
 \frac{1}{i \hbar}(E_{\beta} - E_{\alpha})\langle P_{\alpha \beta  }\rangle +  \frac{1}{i \hbar}\langle [  P_{\alpha \beta} , V ]\rangle
\end{equation}
We restrict ourselves to the second-order in powers of $V$ in calculating the r.h.s. of (\ref{eq49}). 
Finally we obtain the kinetic equations for $\langle P_{\alpha \beta }\rangle$ in
the form~\cite{kuz05}
\begin{equation}\label{eq57}
 \frac{d \langle P_{\alpha \beta }\rangle }{d t} =  \frac{1}{i \hbar}(E_{\beta} - E_{\alpha})\langle P_{\alpha \beta  }\rangle -
 \frac{1}{\hbar^{2}} \int^{0}_{-\infty} dt_{1} e^{\varepsilon t_{1}} 
 \langle \left[[P_{\alpha \beta}, V], V(t_{1}) \right]\rangle_{q}
\end{equation}
The last term of the right-hand side of Eq.(\ref{eq57}) can be called the generalized "collision integral".
Thus,  we can see that the collision term for the system  weakly coupled to the thermal bath has a convenient
form of the double commutator as for the generalized kinetic equations~\cite{pok}  for the system with
small interaction. It should be emphasized  that the assumption about the model form of the Hamiltonian
(\ref{eq42}) is nonessential for the above derivation. We can start again with the Hamiltonian (\ref{eq42}) 
in which we shall not specify the explicit form of $H _{1}$ and $V$. We assume that the state of the
nonequilibrium system is characterized completely by some set of  average values $\langle P_{k}\rangle$ and the
state of the thermal bath by $\langle H _{2}\rangle$. We confine ourselves to such systems for 
which $[H _{1}, P_{k}] =  \sum_{l} c_{kl}P_{l}$. Then we assume that $\langle V \rangle_{q} \simeq 0$, where $\langle \ldots\rangle_{q}$
denotes the statistical average with the quasi-equilibrium statistical operator of the form
\begin{eqnarray}\label{eq58}
\rho_{q}  
 = Q^{-1}_{q} \exp \left(  - \sum_{k} P_{k}F_{k}(t) 
-  \beta H _{2} \right) 
\end{eqnarray}
and $F_{k}(t)$ are the parameters conjugated with $\langle P_{k}\rangle$. Following the method used above in the derivation 
of  equation (\ref{eq57}), we can obtain the generalized kinetic equations for $\langle P_{k}\rangle$ with an accuracy up to
terms which are quadratic in interaction
\begin{equation}\label{eq59}
 \frac{d \langle P_{k}\rangle }{d t} =  \frac{i}{ \hbar} \sum_{l} c_{kl}\langle P_{l}\rangle -
 \frac{1}{\hbar^{2}} \int^{0}_{-\infty} dt_{1} e^{\varepsilon t_{1}} 
 \langle \left[[P_{k}, V], V(t_{1}) \right]\rangle_{q}
\end{equation}
Hence (\ref{eq57}) is fulfilled for the general form of the Hamiltonian of a small system weakly coupled to a
thermal bath.
%
\subsection{ System in Thermal Bath: Rate and Master Equations}\label{nsobm}
In  section  \ref{nsob},  we have described the kinetic equations for $\langle P_{\alpha \beta }\rangle$ in
the general form. Let us write down   equations (\ref{eq57}) in an explicit form. 
We rewrite the kinetic equations for $\langle P_{\alpha \beta }\rangle$ as
\begin{align}\label{eq76}
 \frac{d \langle P_{\alpha \beta }\rangle }{d t} =  \frac{1}{i \hbar}(E_{\beta} - E_{\alpha})\langle P_{\alpha \beta  }\rangle - \nonumber \\
\sum_{\nu} \left( K_{\beta\nu}\langle P_{\alpha \nu }\rangle  + K^{\dag}_{\alpha \nu} \langle P_{\nu \beta}\rangle \right) +
K_{\alpha \beta,\mu \nu} \langle P_{\mu \nu }\rangle
\end{align}
The following notation were used
\begin{eqnarray}
\label{eq74}
 \frac{1}{i\hbar}\sum_{\mu} 
 \int^{ 0}_{ - \infty} dt_{1}  e^{\varepsilon t_{1}}   \langle \Phi_{ \beta \mu} \phi_{\mu \nu}(t_{1})\rangle_{q} = \nonumber \\
\frac{1}{2\pi}  \sum_{\mu}  \int^{ + \infty}_{ - \infty} d\omega
 \frac{J_{\mu \nu, \beta \mu} (\omega)}{\hbar\omega - E_{\gamma} - E_{\delta} + i\varepsilon} = K_{\beta\nu}  \\
 \frac{1}{i\hbar} \int^{ 0}_{ - \infty} dt_{1}  e^{\varepsilon t_{1}} 
(\langle \Phi_{ \mu \alpha} \phi_{\beta \nu}(t_{1})\rangle_{q} + 
\langle \phi_{\mu \alpha}(t_{1}) \Phi_{\beta \nu}\rangle_{q})  = \nonumber \\
\frac{1}{2\pi}   \int^{ + \infty}_{ - \infty} d\omega J_{\beta \nu, \mu \alpha} (\omega)
 \left( \frac{1}{\hbar\omega - E_{\beta} + E_{\nu} + i\varepsilon}  
 - \frac{1}{\hbar\omega - E_{\alpha} - E_{\mu} - i\varepsilon}\right)  \nonumber \\ =  K_{\alpha \beta,\mu \nu} 
\label{eq75}
\end{eqnarray}
Let us now remind~\cite{zub74} that the correlation functions $ \langle A B(t)\rangle$  and  $ \langle A(t) B\rangle$ can be expressed 
via their spectral weight
function ( or spectral intensity) $J(\omega)$
\begin{eqnarray}
\label{eq62} F_{AB} (t-t') = \langle A (t) B(t') \rangle =  \frac {1}{2\pi}
\int^{ + \infty}_{ - \infty} d\omega
\exp [i\omega (t - t')] J_{AB} (\omega)  \\
F_{BA} (t'-t) = \langle B(t') A(t) \rangle =  \frac {1}{2\pi} \int^{ +
\infty}_{ - \infty} d\omega \exp [i\omega (t'-t)] J_{BA} (\omega)
\label{eq63}
\end{eqnarray}
The correlation functions $ \langle \Phi_{ \beta \mu} \phi_{\mu \nu}(t_{1})\rangle_{q}$ 
 and $\langle \phi_{\nu \mu}(t_{1}) \Phi_{\mu \alpha}\rangle_{q}$  are connected with their spectral intensities in the
following way:
\begin{eqnarray}
\label{eq72}
\langle \Phi_{\mu \nu} \phi_{\gamma \delta}(t)\rangle_{q} = \frac {1}{2\pi}\int^{ + \infty}_{ - \infty} d\omega 
J_{\gamma \delta,\mu \nu} (\omega) \exp [-i( \omega - \frac{E_{\gamma} - E_{\delta}}{ \hbar})t] \\
\langle \phi_{ \mu \nu}(t) \Phi_{\gamma \delta}\rangle_{q}  = \frac {1}{2\pi}\int^{ + \infty}_{ - \infty} d\omega 
J_{\gamma \delta,\mu \nu} (\omega) \exp [i( \omega + \frac{E_{\mu} - E_{\nu}}{ \hbar})t] 
\label{eq73}
\end{eqnarray}
The above result is similar in structure to the Redfield equation for the spin density matrix~\cite{red57} when
the external time-dependent field is absent. Indeed, the Redfield equation of motion for the spin density matrix
has the form~\cite{red57}
$$ \frac{  \partial \rho^{\alpha \alpha' }}{\partial t} = - i \omega_{\alpha \alpha'}\rho^{\alpha \alpha' } + 
\sum_{\beta \beta'} R_{\alpha \alpha' \beta \beta'}\rho^{\beta \beta' }.$$
Here $\rho^{\alpha \alpha' } $ is the $\alpha, \alpha'$ matrix element of the spin density matrix,
 $\omega_{\alpha \alpha'} =  (E_{\alpha} - E_{\alpha'}) \hbar  $, where $E_{\alpha}$ is energy of the spin state $\alpha$
and $R_{\alpha \alpha' \beta \beta'}\rho^{\beta \beta' } $ is the "relaxation matrix". A sophisticated analysis and
derivation of the Redfield equation for the density of a spin system immersed in a thermal bath was given in~\cite{cd}. 
A brief discussion of the derivation of the Redfield-type equations in an external field is given in Appendix A.\\
Returning to  Eq.(\ref{eq76}), it is easy to see that if one confines himself to the diagonal averages $\langle P_{\alpha \alpha}\rangle$ only, 
this equation  may be transformed to give
\begin{equation}\label{eq77}
 \frac{d \langle P_{\alpha \alpha }\rangle }{d t} =  \sum_{\nu} K_{\alpha \alpha,\nu \nu} \langle P_{\nu \nu }\rangle
 - \left( K_{\alpha \alpha}  + K^{\dag}_{\alpha \alpha}  \right)\langle P_{\alpha \alpha }\rangle
\end{equation}
\begin{eqnarray}\label{eq78}
K_{\alpha \alpha,\beta \beta} =  \frac{1}{\hbar^{2}} J_{\alpha \beta, \beta \alpha} ( \frac{E_{\alpha} - E_{\beta}}{\hbar} ) = 
W_{\beta \rightarrow \alpha}\\
 K_{\alpha \alpha}  + K^{\dag}_{\alpha \alpha}  = 
 \frac{1}{\hbar^{2}}\sum_{\beta}  J_{\beta \alpha, \alpha \beta} ( \frac{E_{\beta} - E_{\alpha}}{\hbar} ) = 
W_{\alpha \rightarrow \beta} 
\end{eqnarray}
Here $W_{\beta \rightarrow \alpha}$ and $W_{\alpha \rightarrow \beta}$ are the transition probabilities
expressed in the spectral intensity terms. Using the properties of the spectral intensities~\cite{zub74},
it is possible to verify that the transition probabilities satisfy the relation of the detailed
balance
\begin{equation}\label{eq79}
\frac{W_{\beta \rightarrow \alpha}}{W_{\alpha \rightarrow \beta}} = 
\frac{\exp (-\beta E_{\alpha})}{\exp (-\beta E_{\beta})}
\end{equation}
Finally, we have
\begin{equation}\label{eq80}
 \frac{d \langle P_{\alpha \alpha }\rangle }{d t} =  \sum_{\nu} W_{\nu \rightarrow \alpha}  \langle P_{\nu \nu }\rangle
 - \sum_{\nu} W_{\alpha \rightarrow \nu}  \langle P_{\alpha \alpha}\rangle
 \end{equation}
This equation has the usual form of the Pauli master equation. \\ It is well known that  "the master equation
is an ordinary differential equation  describing the {\em reduced evolution} of the system  obtained from the full
Heisenberg evolution by taking the partial expectation with respect to the vacuum state of the reservoirs degrees
of freedom".
The rigorous mathematical derivation of the generalized master equation~\cite{vh,zw,fk} is
rather a complicated mathematical problem. 
%
%
%
\section{DYNAMICS OF NUCLEAR SPIN SYSTEM}\label{dnss}
%
In nuclear magnetic resonance one has a system of nuclei with magnetic moment $\vec{\mu}$ and spins $ \vec{I}$  which are
placed in a magnetic field $h_{0}$. The magnetic moment $\vec{\mu}$ and momentum of nuclei $\vec{J} = \hbar \vec{I}$ are 
related as $\vec{\mu} = \gamma_{n}\vec{J} = \gamma_{n} \hbar \vec{I} = g_{n}\eta \vec{I} $, where $\gamma_{n}$ is the 
gyromagnetic nuclear factor, $g_{n}$ is the nuclear spectroscopic factor, and $\eta  = e\hbar/2Mc $ is the nuclear magneton.
If the spins are otherwise independent, their interaction with the imposed
field produces a set of degenerate energy levels  which for a system of N spins are $(2I + 1)^{N}$ in number  with the energy
spacing $\hbar \omega_{n} = \mu h_{0} /I$. It should be noted that the method of NMR is most powerful and useful in diamagnetic
materials. Metals may be studied, although there are some technical specific problems.\\
In a nuclear-magnetic-resonance experiment, the nuclear spin system absorbs energy from the externally applied 
radio-frequency field and transfers it to the thermal bath or reservoir provided by the lattice through the 
spin-lattice interactions. The latter process requires a time interval of the order of the spin-lattice
relaxation time $T_{1}$. The term "lattice" is used here to denote the equilibrium heat reservoir with temperature $T$
associated with all degrees of freedom of the system other than those associated with the nuclear spins.\\
A great advantage of magnetic resonance method is that the nuclear spin system is only very weakly 
coupled to the other degrees of freedom of the complex system in which it resides and its thermal
capacity is extremely small. It is, therefore,  possible to cause the nuclear spin system itself to depart
severely from thermal equilibrium while leaving the rest of the material essentially in thermal equilibrium.
As a consequence, the disturbance of the system other than the nuclear spins could be ignored. \\
If the nuclei are in thermodynamic equilibrium with the material at temperature $T$ in a field $h_{0}$, a nuclear 
paramagnetic moment $M_{0}$ is produced in the direction of $h_{0}$ given by the Curie formula
$M_{0}/h_{0} = n\mu^{2}/3kT$, $n$ is the number of nuclei per unit volume.\\ We can evidently disturb the system
from equilibrium by applying radiation from outside with quanta of size $\hbar\omega_{n}$ and with suitable polarization. 
If the equilibrium distribution is
disturbed and the population changed, the magnetization in the z-direction, $M_{z}$, is different 
from $M_{0}$, say $M^{h}_{z}$. If then left alone, $M_{z}$ reverts to $M_{0}$ and usually does so exponentially with
time, i.e.
$$M_{z} (t) =  M_{0} -  (M_{0} - M^{h}_{z} ) \exp \{ -\frac{t}{T_{1}} \}$$
The last expression serves to define the spin-lattice relaxation time,  $T_{1}$, and is so called because the process involves exchange of 
magnetic orientation energy with thermal energy of other degrees of freedom ( known conventionally as a lattice ).
All the interactions with the nucleus may contribute to the relaxation process so we must add all contributions to $1/T_{1}$
$$\frac{1}{T_{1}} \propto \frac{1}{T_{1\alpha}} + \frac{1}{T_{1\beta}} + \frac{1}{T_{1\gamma}} + \ldots, $$
where  various contributions to relaxation due to various interactions have been added.
The relaxation rates may be dominated by one or more different physical interactions, so that the observable power 
spectrum may be the Fourier transform of functions involving dipole-dipole correlations, electric field gradient-nuclear
quadrupole moment correlations, etc.\\ The dipole-dipole interaction Hamiltonian $H_{dd}$ between the magnetic moments of  nuclei
may contribute significantly to the nuclear magnetic relaxation process~\cite{hub}.
Consider an explicit 
interaction between the moments $\vec{\mu}_{1}$ and $\vec{\mu}_{2}$ which are distant by $\vec{r}_{12}$ from 
each  other. Then  the interaction is written as
\begin{eqnarray}\label{eq80a}
 H_{dd} = \frac{\vec{\mu}_{1}\vec{\mu}_{2}}{r^{3}_{12}} - 
\frac{3 (\vec{\mu}_{1} \vec{r}_{12})(\vec{\mu}_{2}\vec{r}_{12})}{r^{5}_{12}} = \\ 
- \sqrt{\frac{4\pi}{5}} \frac{1}{r^{3}_{12}} [ 2 \mu^{z}_{1} \mu^{z}_{2} Y_{2,0} - (\mu^{+}_{1} \mu^{-}_{2}  + 
 \mu^{-}_{1} \mu^{+}_{2})Y_{2,0} + \nonumber \\ \sqrt{3} (\mu^{+}_{1} \mu^{z}_{2}   + \mu^{z}_{1} \mu^{+}_{2})Y_{2,-1} + 
\sqrt{3}(\mu^{-}_{1} \mu^{z}_{2}   + \mu^{z}_{1} \mu^{-}_{2})Y_{2,1} \nonumber \\ + \sqrt{6}\mu^{+}_{1} \mu^{+}_{2} Y_{2,-2}
+ \sqrt{6}\mu^{-}_{1} \mu^{-}_{2} Y_{2,2} ] \nonumber
\end{eqnarray} 
where $ \mu^{\pm} = (\mu^{x}  \pm \mu^{y})/ \sqrt{2}$ and $Y_{2,m}$ denote the normalized spherical harmonics of the 
second degree expressed in the form
\begin{eqnarray}\label{eq80b}
Y_{2,\pm 2} = \sqrt{\frac{15}{32\pi}}\sin^{2}\theta_{12} \exp(\pm 2i\phi );\quad Y_{2,\pm 1} = 
\sqrt{\frac{15}{8\pi}}\sin \theta_{12}  \cos \theta_{12} \exp(\pm i\phi ); \nonumber \\
Y_{2,0} = \sqrt{\frac{5}{16\pi}} (3\cos^{2}\theta_{12} - 1 )\nonumber
  \end{eqnarray} 
The dipole-dipole coupling provides the dissipation mechanisms in the spin system. It acts as time dependent perturbations on
the Zeeman energy levels, which results in the relaxation of the nuclear magnetization. \\ Thus, such a spin system can be
described as a superposition of a number of subsystems. They are the Zeeman subsystem for each  spin species
and the dipole-dipole subsystem. A weak applied rf field can be considered as an additional subsystem. The coupling
inside each subsystem is strong,  whereas the coupling between subsystems is weak. As a consequence, the subsystems reach
internal thermal equilibrium independently of each other and one can ascribe a temperature, an energy, an entropy, etc.,
to each of them. Let us note that the usual prediction of statistical mechanics that the temperatures of 
interacting subsystems become equal in equilibrium is a direct consequence of the conjecture that the total energy
is the only analytic constant of the motion.
%
%
\subsection{The Hierarchy  of Time Scales}\label{ts}
%
A case of considerable practical interest in connection with the phenomenon of resonance and relaxation is that of
the hierarchy of time scales. In the standard situations the interaction between nuclear spins is
weak as well as the interaction with the lattice is weak. As a result, in the NMR case the thermal bath variables change on the
fast time scale characterized by $t_{Lc}$ while the spin variables change on the slow time scale characterized by $\tau_{sr}$.\\
First of all, consider the most important concept of spin temperature~\cite{gold}.
Actually, spin systems are never completely isolated and the concept of spin temperature is meaningful only if the
rate $\tau^{-1}_{0}$ of achievement of internal equilibrium is much faster than the spin-lattice relaxation
rate $T^{-1}_{1}$. For time intermediate between $\tau_{0}$ and $\tau_{sr}$, the spin temperature exists and can be 
different from the  lattice temperature $T$. The necessary condition for the applicability of spin temperature concept
is then inequality $\tau_{0} \sim T_{2} \ll T_{1}$.\\
Characteristic times are long in comparison with the time of
achievement of internal equilibrium  in the lattice $t_{Lc}$ but short compared to spin relaxation times
$ t_{Lc} < t < \tau_{sr}$. In this case, the second-order perturbation theory is valid in the weak spin-lattice coupling
parameter. Usually, it is assumed that the time $t_{Lc}$ is very short and $\tau_{0} \geq t_{Lc}$. The restriction of
ordinary perturbation theory generally applied is that it is valid when within the time interval considered the density
matrix cannot change substantially. Argyres and Kelley~\cite{ak} removed the restriction $ t_{Lc} < \tau_{sr}$ and
derived an equation of motion for the spin density that depends on the history of the system~\cite{cd}.\\ One of the
essential virtues of the NSO method is that it focuses attention, at the outset, on the existence of different time scales.
Suppose that the Hamiltonian of the spin system can be divided as $ H = H_{0} + V$, where $ H_{0}$ is the dominant
part, and $V$ is a weak perturbation. The separation of the Hamiltonian into $ H_{0}$ and $V$ is not unique and 
depends on the physical properties of the system under consideration. The choice of the operator $ H_{0}$ determines
a short time scale $\tau_{0}$. This choice is such that for times $t \gg \tau_{0}$ the nonequilibrium state of the 
system can be described with a reasonable accuracy by the average values of some finite set of the operators 
$P_{m}$ ( \ref{eq2}).\\
After the short time $\tau_{0}$, it is supposed that  the system can  achieve the state of an incomplete or 
quasi-equilibrium state.
The main assumption about the  quasi-equilibrium state is that it is determined completely by the quasi-integrals
of motion which are the internal parameters of the system. The characteristic relaxation time of these
internal parameters is much longer than $\tau_{0}$. Clearly then, that even if  these quasi-integrals  at 
the initial moment had no  definitive equilibrium values, after the time $\tau_{0}$, at the quasi-equilibrium state,
those parameters which  altered quickly became the functions of the external parameters and of the quasi-integrals of  
motion. It is essential that this functional connection does  not depend on the initial values of the parameters. In other words,
the operators $P_{m}$ are chosen so that they should satisfy the condition : $[P_{k} , H_{0} ] = \sum_{l} c_{kl}P_{l}$. 
It is necessary to write down the transport equations ( \ref{eq19b}) for this set of "relevant" operators only. The equations
of motion for the average of other  "irrelevant" operators ( other physical variables) will be in some sense   
consequences of these transport equations.
As for the "irrelevant" operators which do not belong to the reduced set of the "relevant" operators $P_{m}$, 
relation $[P_{k} , H_{0} ] = \sum_{l} c_{kl}P_{l}$ leads to the infinite chain of operator equalities.  For times
$t \leq \tau_{0}$ the nonequilibrium averages of these operators oscillate fast, while for  times $t > \tau_{0}$ they become 
functions of the average values of the operators.
%
%
\subsection{Nuclear Spin-Lattice Relaxation}\label{nslr}
%
At the earlier stage, the theory of spin relaxation was developed by means of quantum mechanical 
perturbation methods. Here the spin relaxation is studied by making use of the method of nonequilibrium
statistical operator.
We discuss in this section an arbitrary nuclear spin system on a lattice in interaction with
external fields and another system~\cite{hub}, to be taken eventually to act as a heat bath.
The bath is considered as a quantum-mechanical system  that remained in thermodynamic equilibrium 
while its exchange of energy with the spin system is taken into account.
We consider the processes occurring after  switching off the external
magnetic field in a nuclear spin subsystem of a crystal. Let us consider the behavior of a spin
system with the Hamiltonian $ H_{n}$ weakly coupled by a time-independent perturbation  $V$ to a
thermal bath (temperature reservoir) or a crystal lattice with the Hamiltonian $H_{L}$.\\ The
total Hamiltonian has the form
\begin{equation}\label{eq81}
 H = H_{n} + H_{L} + V,
\end{equation}
where
\begin{equation}\label{eq82}
 H_{n} = - a \sum_{i} I_{i}^{z};  \quad a = \gamma_{n} h_{0}
\end{equation}
Here $I_{i}^{z}$ is the operator of the z-component of the spin at the site $i$ , $h_{0}$ is the time-independent
external field applied in the z-direction, and $\gamma_{n}$ is the gyromagnetic coefficient.\\ Now we 
introduce $b^{\dagger}_{i \lambda}$ and $b_{i \lambda }$ the creation and annihilation operators of the
spin in the site $i$ with the z-component of the spin equal to $\lambda$ , where $- I \leq \lambda \leq  I$.
Then we have 
\begin{equation}\label{eq83}
I_{i}^{z}  = \sum_{\lambda} \lambda  b^{\dagger}_{i \lambda}b_{i \lambda } = 
\sum_{\lambda} \lambda n_{i \lambda }              
\end{equation}
and, consequently,
\begin{equation}\label{eq84}
 H_{n} =  \sum_{i\lambda} E_{\lambda}  n_{i \lambda }; \quad  E_{\lambda} = -a \lambda
\end{equation}
Following section \ref{nsob} we write the Hamiltonian of the interaction as
\begin{equation}\label{eq85}
V = \sum_{i}\sum_{\mu,\nu}\Phi_{i \nu,i\mu}b^{\dagger}_{i\nu}b_{i \mu}, \quad \Phi_{i\nu, i\mu} = 
 \Phi^{\dagger}_{i\mu, i\nu}
\end{equation}
Here $\Phi_{i \nu,i\mu}$ are the operators acting only on the "lattice" variables. 
The term "lattice" is used here to denote the equilibrium heat reservoir with temperature $T$
associated with all degrees of freedom of the system other than those associated with the nuclear spins.
Then, in agreement with Eq.(\ref{eq47}),
we construct the quasi-equilibrium statistical operator
\begin{equation}\label{eq86}
 \rho_{q}  = \rho _{L} \bigotimes \rho_{n},
\end{equation}
where
\begin{eqnarray}\label{eq87}
\rho _{L} = Q^{-1}_{L} e^{- \beta H_{L} }; \quad Q_{L} = Tr \exp (- \beta H_{L})\\
\rho_{n} = Q^{-N}_{n}\exp \left(- \beta_{n}(t)  H_{n}\right); \quad Q_{n} = 
\frac{\sinh\frac{\beta_{n}(t)}{2}a(2I + 1) }{\sinh \frac{\beta_{n}(t)}{2}a}
\end{eqnarray}
Here $\beta_{n}$ is the reciprocal spin temperature and $N$ is the total number of spins in the system.\\
We now turn to  writing down  the kinetic equations for average values $\langle n_{i\lambda}\rangle = 
\langle b^{\dagger}_{i\lambda}b_{i \lambda}\rangle$. 
We use    the kinetic equation  in the    form (\ref{eq80})
\begin{equation}\label{eq88}
 \frac{d \langle n_{i\lambda}\rangle }{d t} =  \sum_{\nu} W_{\nu \rightarrow \lambda}(ii)  \langle n_{i \nu }\rangle
 - \sum_{\nu} W_{\lambda \rightarrow \nu} (ii) \langle n_{i \lambda}\rangle,
 \end{equation}
where
\begin{eqnarray}\label{eq89}
W_{\lambda \rightarrow \nu} (ii) = \frac{1}{\hbar^{2}} J_{\Phi_{i\nu, i\lambda}\Phi_{i\lambda, i\nu}} ( \frac{E_{\nu} - E_{\lambda}}{\hbar}),\nonumber   \\ 
W_{\nu \rightarrow \lambda}(ii) = \frac{1}{\hbar^{2}} J_{\Phi_{i\lambda, i\nu}\Phi_{i\nu, i\lambda}} ( \frac{E_{\lambda} - E_{\nu}}{\hbar})
\end{eqnarray}
It can be shown that
$$\langle n_{i\lambda}\rangle = \langle n_{\lambda}\rangle = Q^{-1}_{n}\exp [- \beta_{n} E_{\lambda}]$$
Then we obtain
\begin{equation}\label{eq90}
 \frac{d \langle n_{\lambda}\rangle }{d t} =  \sum_{\nu} W_{\nu \rightarrow \lambda}  \langle n_{\nu }\rangle
 - \sum_{\nu} W_{\lambda \rightarrow \nu}  \langle n_{\lambda}\rangle
 \end{equation}
where
\begin{equation}\label{eq91}
W_{\lambda \rightarrow \nu}   =    \frac{1}{N}\sum_{i} W_{\lambda \rightarrow \nu} (ii);  \quad   
W_{\nu \rightarrow \lambda}  =    \frac{1}{N}\sum_{i} W_{\nu \rightarrow \lambda}(ii) 
\end{equation}
It is easily seen that
$$W_{\nu \rightarrow \lambda}  =  \exp [\beta ( E_{\nu} - E_{\lambda})  ]  W_{\lambda \rightarrow \nu} $$
Hence, for $\beta_{n}$ we find the equation
\begin{equation}\label{eq92}
\frac{d \beta_{n} }{d t} =  \frac{1}{2} \frac{\sum_{\nu \lambda}(\lambda - \nu) W_{\lambda \rightarrow \nu} 
\left( 1 - \exp [-( \beta -\beta_{n})(E_{\lambda} -  
 E_{\nu} ) ]\right)\exp [- \beta_{n} E_{\lambda}]}{\frac{Q_{n}}{a} \frac{\partial^{2}\ln Q_{n}}{\partial \beta_{n}^{2}}}
\end{equation}
In the derivation of Eq.(\ref{eq92}) we took into account that $\langle I^{z}\rangle = \sum_{\nu }\nu \langle n_{\nu}\rangle$
and
\begin{equation}\label{eq93}
 \frac{d \langle  I^{z}\rangle }{d t} =  - \frac{1}{a} \frac{d \beta_{n} }{d t}\frac{\partial^{2}\ln Q_{n}}{\partial \beta_{n}^{2}} =
 - \frac{1}{a} \frac{d \beta_{n} }{d t} \left(\langle (I^{z})^{2}\rangle  - \langle I^{z}\rangle^{2} \right).
 \end{equation}
In the high-temperature approximation ($\hbar\omega_{n} \ll kT$ )we obtain
\begin{equation}\label{eq94}
\frac{d \beta_{n} }{d t} =  \frac{\beta - \beta_{n}}{T_{1} },
\end{equation}
where $T_{1}$ is the longitudinal time of the spin-lattice relaxation
\begin{equation}\label{eq95}
 \frac{1}{T_{1} } = \frac{1}{2} \frac{\sum_{\nu \lambda}(\lambda - 
 \nu)^{2} W_{\lambda \rightarrow \nu}}{\sum_{\nu }(\nu)^{2}} 
\end{equation}
The above expression is the well-known Gorter relation~\cite{gor,blo,abr,sl}.
%
%
\section{SPIN DIFFUSION OF NUCLEAR MAGNETIC MOMENT }\label{sd}
%
The concept of spin diffusion was invoked by Bloembergen~\cite{bl49} to explain the magnetic relaxation of nuclei
in diamagnetic solids, which is due to the interaction of the nuclear spins with spin of a paramagnetic impurity
ion. This theoretical approach was further developed in many works~\cite{red59,bu65,lo67,abr,khu,khu1}.
In the previous section, we have discussed  a simple calculation of the longitudinal nuclear spin
relaxation time within the NSO approach. Here we shall extend this treatment in order to obtain a more
sophisticated description of the spin dynamics. Let us, therefore, work out a general formula, using these ideas.\\
Consider a  subsystem of interacting nuclear spins $\vec{I}$ of a crystal which interact with the external
magnetic field $h_{0}$ and with other subsystems of a crystal. Our aim is to derive the evolution equation
for the reciprocal spin temperature of the Zeeman spin subsystem $\beta_{n}(\vec{r},t )$ which is relaxed to the equilibrium
after switching off the external rf field.
The total Hamiltonian has the form
\begin{equation}\label{eq96}
 H = H_{n} + H_{dd} + H_{L} + V,
\end{equation}
where the Zeeman operator $H_{n}$ is given by
\begin{equation}\label{eq97}
 H_{n} = -a \sum_{i} I_{i}^{z};  \quad a = \gamma_{n} h_{0}
\end{equation}
It is convenient to rewrite $H_{n}$ in the following form:
\begin{equation}\label{eq98}
\mathcal H_{n}( \vec{r} ) =  \sum_{i} I_{i}^{z}\hbar(\omega_{n} + \Omega_{i})\delta(\vec{r} - \vec{r_{i}} ) 
\end{equation}
Here $\Omega_{i}\ll \omega_{n}$ is  effective renormalization of the "bare" nuclear spin 
energy $\hbar \omega_{n}$ due to the surrounding medium and will be
written explicitly below;
$H_{dd}$ is the operator  of dipole-dipole interaction (\ref{eq80a})
$$H_{dd} = \frac{g_{1} g_{2}\eta^{2}}{r^{3}}\sum_{ij}  \{\vec{I}_{i}\vec{I}_{j} - 
3 (\vec{I}_{i}\hat{r} )(\vec{I}_{j}\hat{r} )\},$$
where $r $ is the distance between the two spins and $\hat{r} = \vec{r}[|\vec{r}|]^{-1}$ is the unit vector in
the direction joining them.
It was shown~\cite{gold,abr} that the so-called secular part of this operator was essential, and
 in the rest of the paper we will use the notation $H_{dd}$ for the secular part of the operator of 
 dipole-dipole interaction. It has the form~\cite{gold,abr}
\begin{equation}\label{eq99}
H_{dd} =  \sum_{i\neq j} A_{ij}\left( I_{i}^{z} I_{j}^{z} - \frac{1}{4} (I_{i}^{+} I_{j}^{-} + I_{i}^{-} I_{j}^{+} ) \right )\\
= \sum_{i\neq j} A_{ij}(I_{i}^{z} I_{j}^{z} - \frac{1}{2} I_{i}^{+} I_{j}^{-})
\end{equation}
Here
$$A_{ij} = \frac{\gamma^{2}_{n}\hbar}{2r^{3}_{ij}}( 1 - 3\cos^{2} \theta_{ij})$$
and $\theta_{ij}$ is the angle between $\vec{h_{0}}$ and $\vec{r_{ij}}$;\\
$H_{L}$ is the Hamiltonian of the a thermal bath and  $V$ is the operator of interaction between the nuclear 
spins and the lattice.
Since our aim is to derive the equation for the relaxation of the Zeeman energy, we take the operators $\mathcal H_{n}( \vec{r} )$ and $H_{dd}$ 
as the relevant variables which describe the nonequilibrium state.  According to the
NSO formalism, we now write the entropy operator  (\ref{eq44})     in the form
\begin{eqnarray}\label{eq100}
S(t,0) = \Omega(t) + \beta H_{L} +  \beta_{d} H_{dd} + \int \beta_{n}(\vec{r},t )\mathcal H_{n}( \vec{r} )d^{3}r, \\ \nonumber
\rho_{q} (t) =  \exp (- S(t,0))
\end{eqnarray}
where $\beta_{d}$ and $\beta$ are the reciprocal temperature of dipole-dipole subsystem and the thermal bath,
respectively. Then, within the formalism of NSO, as described above in section  \ref{nsob}, it is possible to
derive the corresponding transport equations for the nonequilibrium averages $\langle \mathcal H_{n}( \vec{r} )\rangle$ and
$\langle H_{dd}\rangle$. Here we confine ourselves to the equation for the  $\langle \mathcal H_{n}( \vec{r} )\rangle$ since the
equations for $\beta_{n}(\vec{r},t )$ and $\beta_{d}$ are decoupled when the external rf field is equal to zero.
We need the relations
\begin{eqnarray}\label{eq100a}
\frac{d \mathcal H_{n}( \vec{r} )}{dt} = \frac{1}{i\hbar}[\mathcal H_{n}( \vec{r} ),V ] + 
\frac{1}{i\hbar}[\mathcal H_{n}( \vec{r} ),  H_{dd}] = 
K_{n} ( \vec{r} )  - div \vec{J}( \vec{r} )
\end{eqnarray}
Here $K_{n} ( \vec{r} )$ is the source term and $\vec{J}( \vec{r} )$ is the effective nuclear spin energy current
\begin{equation}\label{eq100b}
\vec{J}( \vec{r} )  =  \frac{1}{2i}\sum_{k\neq l} A_{kl}\vec{r}_{kl}(\omega_{n} + \Omega_{l})\delta(\vec{r} - \vec{r_{k}} )
I_{k}^{+} I_{l}^{-}   
\end{equation}
Since $\Omega_{i}\ll \omega_{n}$, the approximate form of the current is
\begin{equation}\label{eq100c}
\vec{J}( \vec{r} )  \approx  \frac{\omega_{n}}{2i}\sum_{k\neq l} A_{kl}\vec{r}_{kl} \delta(\vec{r} - \vec{r_{k}} )
I_{k}^{+} I_{l}^{-}   
\end{equation}
The law of conservation of energy in the differential form can be written as (c.f.~\cite{bu65})
\begin{equation}\label{eq100d}
\frac{d \langle \mathcal H_{n}( \vec{r} ) \rangle}{dt} =   - div  \langle \vec{J}( \vec{r} ) \rangle + 
 \langle K_{n} ( \vec{r} )\rangle
\end{equation}
Following  the method of calculation of Buishvili and Zubarev~\cite{bu65}  we get
\begin{eqnarray}\label{eq101}
\frac{\partial \langle \mathcal H_{n}( \vec{r} )\rangle  }{\partial t} =
 - \sum_{\mu\nu=1,2,3} \frac{\partial }{\partial x_{\mu}} L^{\mu\nu}( \vec{r} )\frac{\partial }{\partial x_{\nu}}\beta_{n}(\vec{r},t ) +
(\beta_{n}(\vec{r},t ) - \beta )L_{1}( \vec{r} )
\end{eqnarray}
According to Eq.( \ref{eq98}), we have treated $ \langle I^{z} (t )\rangle$ as a continuum function of spatial
variables so that when evaluated at the lattice site $j $, it is equal to $ \langle I^{z}_{j} (t )\rangle$. Carrying out
a Taylor series expansion~\cite{hanh} of $ \langle I^{z} (t )\rangle$ about  the $k$th lattice site and then evaluating the results
at position $j$ yield 
\begin{eqnarray}\label{eq101a}
\langle I^{z}_{j} (t )\rangle \approx \langle I^{z}_{k} (t )\rangle + 
\sum^{3}_{\alpha=1} \frac{\partial}{\partial x^{\alpha}}  \langle I^{z} (t )\rangle|_{k} x_{kj^{\alpha}} +  \\ \nonumber
\frac{1}{2}\sum^{3}_{\alpha,\beta=1} 
\frac{\partial^{2}}{\partial x^{\alpha}\partial x^{\beta} }  \langle I^{z} (t )\rangle|_{k} x_{kj^{\alpha}}x_{kj^{\beta}} 
+ \hdots ,
\end{eqnarray}
where $x_{kj^{\alpha}}$ is the $\alpha$ coordinate ($\alpha = 1, 2,3$ ) in an arbitrary Cartesian coordinate system for
$ \vec{r}_{kj}$, and $\partial / \partial x^{\alpha} \langle I^{z} (t )\rangle|_{k}$ is the partial derivative of 
$ \langle I^{z} (t )\rangle$ with respect to $x^{\alpha}$, evaluated at the lattice site $k$.\\
The generalized kinetic coefficients $L^{\mu\nu}( \vec{r} )$ and $L_{1}( \vec{r} )$ have the form
%
%
\begin{align}
\label{eq102}
L^{\mu\nu}( \vec{r} ) = \int^{0}_{- \infty}dt_{1}\int^{1}_{0}d\lambda \int d^{3}q
\langle J_{\mu}( \vec{r} )\exp ( - \lambda S(t,0)) J_{\nu}( \vec{q},t_{1} ) \exp ( \lambda S(t,0))\rangle_{q} \\
L^{1}( \vec{r} ) = \int^{0}_{- \infty}dt_{1}\int^{1}_{0}d\lambda \int d^{3}q
\langle K_{n}( \vec{r} )\exp ( - \lambda S(t,0))K_{n}( \vec{q},t_{1}  ) \exp ( \lambda S(t,0))\rangle_{q}
\label{eq103}
%
\end{align}
The condition $\langle \mathcal H_{n}( \vec{r} )\rangle = \langle \mathcal H_{n}( \vec{r} )\rangle_{q} $ determines the connection of
$\beta_{n}(\vec{r},t )$ and $\langle \mathcal H_{n}( \vec{r} )\rangle $. Equation ( \ref{eq101}) is the 
diffusion type equation~\cite{mf,ar,sta}. This equation describes more fully the local changes of the Zeeman energy
due to the relaxation and transport processes in the system with the Hamiltonian (\ref{eq96}). 
In its general
form   equation ( \ref{eq101}) is very complicated~\cite{mf,ar,sta} and to get a solution,   various approximate schemes should
be used.
%
\subsection{Evaluation of  Spin Diffusion Coefficient}\label{nsdc}
%
Let us consider the calculation of the diffusion coefficient. The most obvious approximation to express the average
$\langle \mathcal H_{n}( \vec{r} )\rangle$ in terms of $\beta_{n}(\vec{r},t )$ is the high-temperature 
approximation $\beta F_{n}(t) \ll 1$ or
$\hbar\omega_{n} \ll kT$. As a rule, this approximation is well fulfilled in the NMR experiment. 
Making  use of
high-temperature expansion in Eq.(\ref{eq101})  and taking into account that in this approximation 
$$
\exp (- S(t,0)) \approx \frac{1}{Tr_{I}1} \left( 1 - 
\int d^{3}r \beta_{n}(\vec{r},t )\mathcal H_{n}( \vec{r} ) \right) \rho_{L}$$
we get
\begin{eqnarray}\label{eq105}
\frac{\partial \beta_{n}(\vec{r})  }{\partial t} =
 \sum_{\mu\nu} \frac{\partial }{\partial x_{\mu}} D^{\mu\nu}( \vec{r} )\frac{\partial }{\partial x_{\nu}}\beta_{n}(\vec{r},t ) -
(\beta_{n}(\vec{r}) - \beta )R( \vec{r} )
\end{eqnarray}
or in a different form
\begin{eqnarray}\label{eq105a}
\frac{\partial \beta_{n}(\vec{r})  }{\partial t} =
 D( \vec{r} )\Delta \beta_{n}(\vec{r}  ) -
(\beta_{n}(\vec{r}) - \beta )R( \vec{r} ).
\end{eqnarray}
Here $D( \vec{r} )$ is the diffusion coefficient 
\begin{eqnarray}\label{eq106a}
D(\vec{r}) = -\frac{1}{2\hbar^{2} \omega_{n}^{2} N(r)} \int^{0}_{- \infty} e^{\varepsilon t_{1}}dt_{1} \int d^{3}r_{1}
\frac{Tr_{I} \langle J( \vec{r} ) J( \vec{r_{1}},t_{1} ) \rangle_{L}} {Tr_{I}( I^{z})^{2}} 
\end{eqnarray}
$N(r) = \sum_{k}\delta (\vec{r} - \vec{r}_{k} )$ being the nuclear spin density.
The quantity $R( \vec{r} ) > 0$  has the following form:  
\begin{eqnarray}\label{eq106}
R(\vec{r}) = -\frac{1}{\hbar^{2} \omega_{n}^{2} N(r)} \int^{0}_{- \infty} e^{\varepsilon t_{1}}dt_{1} \int d^{3}r_{1}
\frac{Tr_{I} \langle K_{n}( \vec{r} ) K_{n}( \vec{r_{1}},t_{1} ) \rangle_{L}} {Tr_{I}( I^{z})^{2}} 
\end{eqnarray}
Here the symbol  $\langle  \hdots \rangle_{L} = Tr (   \hdots \rho_{L})$  implies the average over the
equilibrium ensemble for lattice degrees of freedom.
\subsection{Host Nuclear Spin Diffusion  in Dilute Alloys}\label{hnslr}
%
Spin diffusion is the transport of Zeeman energy or magnetization via the dipole-dipole interactions and it proved important both 
theoretically~\cite{bl49,red59,bu65,bowa,bug69} and experimentally~\cite{red59,blum} in diamagnetic solids.
We consider here another class of substances, the dilute alloys~\cite{rig,hir,pet}.
The spin dynamics and relaxation of bulk metal nuclei by relatively dilute local moments in dilute alloys (e.g. $Cu-Mn$)
was studied quite extensively, both theoretically~\cite{mor,koh,walke,gih,js,walk,gi,zit,noz} and 
experimentally~\cite{br1,br,row,stre,clo,ba,lum,gos,wals,wal,aa,lo,all,allo,bern,wawa,yaf}.
The description of spin relaxation in dilute alloys has certain specific features as compared with the
homogeneous systems. For  brevity we confine ourselves to the consideration of the bulk metal nuclei relaxation
in dilute alloy. Due to the dipole-dipole interaction between a nuclear spin and an impurity spin, the relaxation rate
may become nonuniform. It is more rapid for the spins that are close  to impurity and is much slower for 
the distant nuclear spins. As a result, a nonuniform distribution in the bulk nuclear spin subsystem will occur and
to describe spin relaxation consistently, the nuclear spin diffusion should be taken into account.\\
The Hamiltonian for  nuclear  and electronic  interacting spin subsystems is
\begin{equation}\label{eq107}
H = H_{n} + H_{e} +   H_{M} +  H_{ne} +  H_{Me}  +  H_{dip}
\end{equation}
Here index $n$ denotes the host nuclear spins,  $M$ denotes spin of the magnetic impurities, and $e$ denotes the
electron subsystem. In this section, when we refer to the host nuclear spin subsystem $H_{n}$ we put
\begin{equation}\label{eq108}
H_{n} =   \sum_{i} I_{i}^{z}\hbar \omega_{n}  +
\sum_{i\neq j} A_{ij}(I_{i}^{z} I_{j}^{z} - \frac{1}{2} I_{i}^{+} I_{j}^{-}). 
\end{equation}
The Hamiltonian of electron subsystem is
\begin{equation}\label{eq109}
 H_{e} = \sum_{k\sigma}\varepsilon_{k\sigma}a^{\dag}_{k\sigma}a_{k\sigma}
\end{equation}
and
\begin{equation}\label{eq110}
 H_{M} = \sum_{m}\hbar \omega_{M}S^{z}_{m}
\end{equation}
is the Hamiltonian of the impurity spins in the external magnetic field. The Hamiltonian of the interaction~\cite{yaf} of
nuclear spins and the spin density $ \vec{\sigma}(\vec{R_{i}}) $ of the conduction electrons is
\begin{equation}\label{eq111}
 H_{ne} = J_{ne}\sum_{i} \vec{I_{i}}\vec{\sigma}(\vec{R_{i}}); \quad  J_{ne} = 
 - \frac{8\pi}{\hbar^{2}\gamma_{n} \gamma_{e}},  
\end{equation} 
where
$$ \sigma^{+}_{k} = \sum_{q} a^{\dag}_{q\uparrow}a_{k+q\downarrow},
\quad \sigma^{-}_{-k} = ( \sigma^{+}_{k} )^{\dagger} = \sum_{q}
a^{\dag}_{k+q\downarrow}a_{q\uparrow}$$
Interaction of the impurity spins $\vec{S}_{m}$ and the spin density of the itinerant carriers is given by the
spin-fermion~\cite{rk,kuze05}  ($sp-d(f)$)model Hamiltonian
\begin{equation}\label{eq112}
 H_{Me} = J_{sd}\sum_{m} \vec{S_{m}}\vec{\sigma}(\vec{R_{m}})
\end{equation} 
The last part of the total Hamiltonian (\ref{eq107})
\begin{equation}\label{eq113}
 H_{dip} = \hbar \sum_{im}\sum_{\mu\nu = x,y,z} \Phi^{\mu\nu}_{im}I^{\mu}_{i}S^{\nu}_{m},
\end{equation}
is the Hamiltonian of the dipole-dipole and pseudo-dipole interaction of nuclear and impurity spins. This interaction
was described in detail in Refs.~\cite{br1,br,vv}. 
The pseudo-dipolar interaction does not  originate in crystalline anisotropy  but in the tensor
character of the dipolar interaction~\cite{br}. Their expression for the pseudo-dipolar interaction is
\begin{equation}\label{eq114}
 H^{PD}_{nn} = \sum_{ij} [ \vec{I}_{i} \vec{I}_{j} - 3r_{ij}^{-2} (\vec{I}_{i}\vec{r_{ij}} )(\vec{I}_{j}\vec{r_{ij}} ) ]B_{ij}
\end{equation}
The Van Vleck Hamiltonian for a system with two magnetic ingredients~\cite{br,vv}
includes the term
\begin{eqnarray}\label{eq115}
H_{dip} = \sum_{i>j} (\frac{ g_{n}^{2}\eta^{2}}{r^{3}_{ij}} + \tilde{B}_{ij})   \{\vec{I}_{i}\vec{I}_{j} - 
3r^{-2}_{ij} (\vec{I}_{i}\vec{r_{ij}} )(\vec{I}_{j}\vec{r_{ij}} )\} \nonumber \\ +
\sum_{im} (\frac{g_{n} g_{e}\eta^{2}}{r^{3}_{im}} + \tilde{B}_{im}) \{\vec{I}_{i}\vec{S}_{m} - 
3r^{-2}_{ij} (\vec{I}_{i}\vec{r_{im}} )(\vec{S}_{m}\vec{r_{im}} )\} \nonumber \\ +
\sum_{m>n} (\frac{ g_{e}^{2}\eta^{2}}{r^{3}_{mn}} + \tilde{B}_{mn})   \{\vec{S}_{m}\vec{S}_{n} - 
3r^{-2}_{mn} (\vec{S}_{i}\vec{r_{mn}} )(\vec{S}_{j}\vec{r_{mn}} )\}.
\end{eqnarray}
The $\tilde{B}$'s represent the pseudo-dipolar interaction
$$B_{ij} = \frac{3}{2}(\tilde{B}_{ij} + \frac{ g_{n}^{2}\eta^{2}}{r^{3}_{ij}})
( 1 - 3\cos^{2} \theta_{ij})$$.
The later consists of three components of which we use in Eq.(\ref{eq113}) the following one as the most
essential~\cite{vv,br} 
\begin{equation}\label{eq116}
 H^{PD}_{Mn} = \sum_{im}B_{im}\vec{I}_{i} (\vec{S}_{m} - \hat{r}_{im}(\hat{r}_{im} \vec{S}_{m} ))
\end{equation}
It was shown in~\cite{br} that for the large distance between the nuclear spin and the electron spin $B_{im}$ has the form
\begin{equation}\label{eq117}
B_{im} \approx B \frac{\cos (k_{F}r_{im} + \phi_{B} )}{(2k_{F}r_{im})^{3}}
\end{equation}
Thus, in structure, the coefficient $B_{im}$ is similar to the production of the contact potential and the spatial part
of the RKKY interaction~\cite{coq}. As a rule, the pseudo-dipolar interaction is less than the contact interaction. The estimations
give $ B \sim 1/3 J_{ne}$ for $^{205}Tl$. It will be even more valid for copper since its mass is much less than for $Tl$.\\
Now the expression for the Hamiltonian $H_{dip}$ can be rewritten as
\begin{eqnarray}\label{eq118}
H_{dip} = \gamma_{n}\gamma_{M}\hbar\sum_{im}\frac{1}{r_{im}^{3}}\{ I_{i}^{z}\delta S_{m}^{z}( 1 - 3\cos^{2} \theta_{im}) - \\
\frac{3}{2} \sin \theta_{im}\cos \theta_{im} [\exp( - i\phi_{im}) I_{i}^{+} \delta S_{m}^{z} + 
\exp(  i\phi_{im}) I_{i}^{-} \delta S_{m}^{z}] \}  \nonumber \\
\{1 + B \frac{\cos (2k_{F}r_{im} + \phi_{B} )}{8k_{F}^{3}}\}  \nonumber
\end{eqnarray}
Here we have introduced the mean field $\langle S_{m}^{z} \rangle$ and the fluctuating part of the impurity spin, namely 
$\delta S_{m}^{z} = S_{m}^{z} - \langle S_{m}^{z} \rangle$. By substituting this definition 
of $S_{m}^{z}$ into (\ref{eq111}) rewritten in terms of the variable $\delta S_{m}^{z}$ we obtain
\begin{eqnarray}\label{eq119}
 H_{ne} = - \frac{8\pi}{\hbar^{2}\gamma_{n} \gamma_{e}}\sum_{ip} (\vec{I_{i}}\vec{\sigma}_{p}) 
\delta(\vec{R_{i}} - \vec{r_{p}}) = \\
J_{ne} \sum_{ip} [(I_{i}^{+} \sigma_{p}^{-} + I_{i}^{-}\sigma_{p}^{+})\delta(\vec{R_{i}} - \vec{r_{p}}) + \nonumber \\
( \sigma_{p}^{z}\delta(\vec{R_{i}} - \vec{r_{p}})  - \langle \sigma_{p}^{z}  
\delta(\vec{R_{i}} - \vec{r_{p}}) \rangle ) I_{i}^{z} ], \nonumber
\end{eqnarray} 
where
$$ \sum_{p}\vec{\sigma}(\vec{r} _{p})\delta(\vec{R_{i}} - \vec{r_{p}}) = 
\sum_{kk'} \sum_{ss'} \langle s|\vec{\sigma}|s'\rangle \psi^{*}_{k'}(0) \psi_{k}(0) a^{\dag}_{ks}a_{k's'}. $$
Now it is possible to write down explicitly the shift of the Zeeman frequency $\omega_{n}$ in  (\ref{eq98}) due to the
mean-field renormalization $\Omega_{i}$ as
\begin{eqnarray}\label{eq120}
 \Omega_{i} = \gamma_{n} \gamma_{M} \hbar \sum_{m} \frac{1}{r_{im}^{3}}
  \{ 1 + B \frac{\cos (2k_{F}r_{im} + \phi_{B} )}{8k_{F}^{3}} \}\langle S^{z} \rangle - \nonumber \\
 J_{ne}\sum_{p} \langle \sigma_{p}^{z}  
\delta(\vec{R_{i}} - \vec{r_{p}}) \rangle   = \sum_{m} \Phi^{zz}_{im} \langle S^{z}_{m} \rangle - 
J_{ne}\sum_{p} \langle \sigma_{p}^{z} \delta(\vec{R_{i}} - \vec{r_{p}}) \rangle 
\end{eqnarray}
This shift of the Zeeman frequency  $(\Omega_{i} \ll \omega_{n} ) $  is the most essential for the evaluation
of the coefficient of spin diffusion~\cite{abr,ab,khu}.
%
\subsection{Spin Diffusion Coefficient in Dilute Alloys}\label{sdcda}
%
%
Here we evaluate concrete expressions for the spin diffusion coefficient (\ref{eq106a}) for the dilute alloys system
which is described by the Hamiltonian (\ref{eq107}). Consider again the approximate equation (\ref{eq105}) where the
diffusion coefficient can be written as
\begin{equation}\label{eq121}
 D^{\mu\nu} \approx \frac{\omega_{d}}{\hbar^{2}\sqrt{\pi}} \sum_{l} A^{2}_{rl} (r^{\mu} - r^{\mu}_{l})
 (r^{\nu} - r^{\nu}_{l}) \exp [- (\Omega_{r} - \Omega_{l} )^{2}/4\ (\omega_{d})^{2}]
\end{equation}
In the derivation of the above expression, to permit explicit calculations, the Gaussian approximation for the
nuclear spin correlation function was used ( see Appendix B). From  equations (\ref{eq105}) and (\ref{eq121}) it 
follows that in the process of the longitudinal nuclear spin relaxation, which is a function of position,  there is a possibility 
to transport  the nuclear magnetization ( i.e. excess of nuclear spin density) due to the dipole-dipole interaction. It
is clearly seen that the nuclei themselves do not move in the spin diffusion process. There is diffusion of the excess of
the projection of the nuclear spin only.\\ To proceed further, consider the case when the concentration of the
impurity spins is very low. In this case, for one impurity spin there is a  big number of   host nuclear spins 
which interact  with it. In other words, this case corresponds to the effective single-impurity situation. Thus,  we can 
place one impurity spin to the origin of the coordinate frame (0,0,0). The  vector $\vec{r}$ in Eq.(\ref{eq121}) is then
counted from  this position. For a simple cubic crystalline system with the inversion center the symmetric tensor
$D^{\mu\nu}(\vec{r})$ is reduced to the scalar $D(\vec{r})$. The coefficient $D(\vec{r})$ decreases with decreasing the 
distance $r$ when $r$ is small. This is related with the fact that Zeeman nuclear frequencies of the nuclei, which are close to the
impurity, have substantially different values due to the influence of the local magnetic fields induced by the impurity
spin. This circumstance hinders the flip-flop $(\Omega = \omega_{M} - \omega_{n})$ transitions of neighboring nuclei since this transition does not
conserve the total Zeeman energy of nuclear spins. ( Let us remind that if we suppose  that the spins $S$ are completely
polarized and the nuclear spins $I$ are completely unpolarized, then the dipolar interaction permits simultaneous
reversals of $S$ and $I$ in the opposite directions, or flip-flops, and also reversals in the same direction  which is usually
called flip-flips with $\Omega = \omega_{M} + \omega_{n}$). In  expression (\ref{eq121}) this tendency is described
by the exponential factor. This exponential factor  leads to the appearance of the so-called "diffusion barrier" around
each impurity. Inside this diffusion barrier the diffusion of nuclear spin is hindered strongly~\cite{abr,khu}.\\
It can be seen that for the large distance from the impurity the frequency difference  in
Eq.(\ref{eq121}) behaves as $ (\Omega_{r} - \Omega_{l}) \ll \omega_{d}$, 
where $\omega_{d} \approx 6 \gamma_{n}^{2} \hbar a^{-3}$ is the dipolar line-width  and $D(r)$ does not depend on $r$. In the opposite case,
of small distance scale ( near impurity ) the frequency difference is big and the coefficient $D(r)$  decreases
quickly with  the distance to the impurity. Thus, it is convenient to introduce the effective radius of the
diffusion barrier $\delta$, namely, a distance from the impurity for which the following definition holds:
\begin{equation}\label{eq122}
 D(r) = 
\begin{cases}
 D & \text{if $r > \delta$}\\  
0 & \text{if $r < \delta$} 
\end{cases} 
\end{equation}
The constant $D$ is equal to $ D = \omega_{d}/3\hbar^{2}\sqrt{\pi} \sum A^{2}_{kl}r^{2}_{kl}$.\\
Let us estimate the "size" of the diffusion barrier. Consider two neighboring nuclei which take up a position  along the radius
from the impurity. The distance between them is equal to the lattice constant $a$. In this case, the frequency shift
is equal to $(\Omega_{\delta} - \Omega_{\delta + a}) \approx \omega_{d}$ 
and  $\delta \approx a \sqrt[4]{[\gamma_{M}/\gamma_{n}\langle S^{z} \rangle ]} $ ( see Appendix B).\\
Consider again the approximate equation (\ref{eq105a}) taking into account the
diffusion barrier approximation (\ref{eq122}). It  can be rewritten in the form
\begin{eqnarray}\label{eq123}
\frac{\partial \beta_{n}(\vec{r},t)  }{\partial t} =
 D\Delta \beta_{n}(\vec{r},t  ) -
(\beta_{n}(\vec{r},t) - \beta ) (R_{0} + R_{1}( \vec{r} ) + R_{2}( \vec{r} )),
\end{eqnarray}
where
\begin{eqnarray}\label{eq124}
R_{0} = \frac{2J_{ne}^{2}}{\hbar^{2}2\pi}\sum_{kk'}\sum_{pp'} \psi^{*}_{k } \psi_{k'}\psi^{*}_{p } \psi_{p'}
\int^{\infty}_{- \infty} d \omega f(\omega - \omega_{n} ) G^{0}_{kk'pp'}(\omega),  \\
G^{0}_{kk'pp'}(\omega) = \int^{\infty}_{- \infty} dt \exp (it \omega) \langle a^{\dag}_{k\downarrow}a_{k'\uparrow}
a^{\dag}_{p\uparrow} (t)a_{p'\downarrow}(t) \rangle,
\end{eqnarray}\label{eq125}
\begin{eqnarray}\label{eq126}
R_{1} ( \vec{r} ) = - \frac{4J_{ne}}{\hbar 2\pi}\sum_{kk'}\sum_{m} 
\int^{\infty}_{- \infty} d \omega f(\omega - \omega_{n} ) Re (\psi^{*}_{k } \psi_{k'} G^{1}_{kk'm}(\omega) \Phi^{+z}_{rm}), \\
G^{1}_{kk'm}(\omega) = \int^{\infty}_{- \infty} dt \exp (it \omega) \langle a^{\dag}_{k \uparrow}a_{k'\downarrow}
S^{z}_{m}(t) \rangle,
\end{eqnarray}\label{eq127}
and
\begin{eqnarray}\label{eq128}
R_{2} ( \vec{r} ) = \frac{9}{2 (\gamma_{n} \gamma_{M}\hbar)^{2}} \frac{1}{2\pi}\sum_{m}
\int^{\infty}_{- \infty} d \omega f(\omega - \omega_{n} )  G^{i}_{mm}(\omega) Y_{m}, \\
G^{i}_{mm}(\omega) = \int^{\infty}_{- \infty} dt \exp (it \omega) \langle \delta S^{z}_{m}
\delta S^{z}_{m}(t) \rangle, \\
Y_{m} = \{ 1 + \frac{B \cos (2k_{F}|\vec{r} - \vec{r}_{m}| + \phi_{B} ) }{8 k_{F}^{3}} \}^{2}
\frac{\sin^{2}\theta_{rm} \cos^{2}\theta_{rm}}{|\vec{r} - \vec{r}_{m}|^{6}}.
\end{eqnarray}\label{eq129}
Here the function  $f(\omega - \omega_{n} )$  is the  NMR line-shape. The line-shape of the NMR spectrum~\cite{mem} arises from the variation of the
local field at a given nucleus because of the interaction with nearby neighbors. The inhomogeneity of the applied magnetic field may also
increase the width of the line. 
\\ The contribution of the factor $R^{-1}_{0}$ leads to the
generalized  Korringa relaxation rate~\cite{kor}
\begin{equation}\label{eq129a}
\frac{1}{T_{1}}  \propto \frac{\pi kT}{\hbar} [ \frac{8\pi}{3}\gamma_{n}\hbar \chi_{p}\frac{M }{\mu_{e}}
\langle |\psi_{F}(0)|^{2}\rangle ]^{2} 
\end{equation}
Korringa~\cite{kor}  calculated the spin-lattice relaxation time $T_{1}$ in metals and  showed that $T_{1}$
should be inversely proportional to temperature and should be related to the Knight shift ( see also Ref.~\cite{tmor}).
Korringa nuclear spin-lattice relaxation occurs in a metal through the nucleus-electron interaction of 
contact type~\cite{kor}
\begin{equation}\label{eq129b}
\frac{8\pi}{3}(|\gamma_{e}|\hbar \vec{s})(\gamma_{n}\hbar \vec{I}) |\psi_{A}(0)|^{2}.
\end{equation}
The quantity $R_{1}$ is determined by the correlation of the electron and impurity spins and is highly anisotropic.\\
The quantity $R_{2}$ is related to the scattering of nuclear spins on the fluctuations of  impurity spins. The last
contribution is the most essential factor in the present context. This is related to the fact that the main characteristic 
features of the problem under consideration clearly manifest itself in the isotropic case which is considered in the
majority of works. In the isotropic case $R_{1} = 0$ and  the contribution of  $R_{2}$ can be expressed as
\begin{eqnarray}\label{eq130}
R_{2} (r) = \sum_{m} C \{1 + \frac{B \cos (2k_{F}|\vec{r} - \vec{r}_{m}| + \phi_{B} ) }{8 k_{F}^{3}} \}^{2}
\frac{1}{|\vec{r} - \vec{r}_{m}|^{6}}\\
C =\frac{3}{5 (\gamma_{n} \gamma_{M}\hbar)^{2}} \frac{1}{2\pi} 
\int^{\infty}_{- \infty} d \omega f(\omega - \omega_{n} )  G^{i}_{mm}(\omega)
\end{eqnarray}\label{eq131}
Nevertheless, even after  simplifications described above, a solution of the diffusion equation is still a
complicated problem. The main difficulty is the presence of the highly oscillating 
factor $ \cos (2k_{F}|\vec{r} - \vec{r}_{m}| + \phi_{B} )$. The role of this oscillating 
factor can be taken into account entirely by numerical calculations. For a qualitative rough estimation  we consider the
simplified case when $B \approx 0$. Then we can proceed following the method of calculation of Ref.~\cite{khu}.
According to these calculations~\cite{khu} we find
\begin{equation}\label{eq132}
\frac{1}{T_{1}} = (R_{0})^{-1}  + 4\pi D N F
\end{equation}
Here $N$ is the number of impurities and the quantity $F$ has the form
\begin{equation}\label{eq133}
 F = 
\begin{cases}
0.7 b & \text{if $b > \delta$}\\  
1/3 (b/\delta)^{3}b & \text{if $b < \delta$} 
\end{cases} 
\end{equation}
where $b = \sqrt[4]{(C/D)}$.\\
It is clear from  Eqs.(\ref{eq132}) and (\ref{eq133}) that the behavior of the relaxation time and its
value depend  strongly on the interrelation of $b$ which is determined by the correlation function
$G^{i}_{mm}(\omega)$ and of $\delta$ which is determined by $\langle S^{z} \rangle$, as well as on the temperature
for  each concrete alloy. Thus, the problem of description of spin-lattice relaxation in dilute metallic alloys was
reduced to the problem of calculation of the value of $F$. When $\delta \ll b$ the diffusion barrier is nonessential.
In the opposite case, when $ b  < \delta $, the  diffusion barrier is essential and leads to 
the slowing down of the relaxation process. In other words, the distance $b$ determines the scale up to which 
the nuclear spin relaxation is effective. Finally, let us note that the order of value of time which is
necessary to transmit the magnetic moment to the distance $r$ in a solid is equal to $ \tau_{D} \simeq r^{2}/D$; 
for $r = 10^{-6}$ cm it gives the value $ \tau_{D} \simeq 1$ sec.
%
%
\section{CONCLUDING REMARKS}
In the present paper, we have given a complementary method for obtaining the rate and relaxation equations of
nuclear spin system in solids. The main tool in this approach is the use of the method of nonequilibrium statistical 
operator~\cite{zub74}. We have presented a theory of spin relaxation  which allows us to derive   general 
equations of spin dynamics. In addition, our theory allows us to take into account the effects of spin diffusion in a 
very straightforward manner. The calculations were kept general by restricting the form of spin-lattice Hamiltonian 
as little as possible.  It has permitted us to perform the
derivation under more general conditions and explicitly demonstrate some key features of irreversible processes
in solids. \\ It was shown that the spin systems provide a  useful proving ground for applying the sophisticated methods
of statistical thermodynamics. The method used is capable of systematic improvement and gives a deeper insight 
into the meaning of the spin relaxation processes in solids. We have shown that the transport of nuclear 
spin energy in a lattice of paramagnetic spins  with magnetic dipolar interaction plays an important role 
in relaxation processes in solids.
To test the general formalism presented here, an example of a dilute metallic alloy system was considered to 
demonstrate the usefulness of the equations derived. \\
In summary, the present paper examines the relaxation dynamics of a spin system. It continues the
investigation presented in the previous work into the use of statistical mechanical methods for systems that are
in  contact with a thermal bath. We used the method of the nonequilibrium statistical operator developed
by D. N. Zubarev.  In the present paper, we have developed the application of this method to the spin-relaxation
problem, so that some useful results may be obtained from it. The calculation presented in this paper can be said to show
 that the NSO method has provided a compact and efficient tool for description of the spin relaxation dynamics. 
In this respect, the present 
treatment may be regarded as a complement of the  Buishvili and Zubarev~\cite{bu65} seminal treatment.\\
Though the analysis of this paper concentrates on the nuclear spin systems in solids, the extension to other
spin systems, e.g., paramagnetic electron spin system, is straightforward. The other important task is to examine the
effects of a periodically time-dependent field on the long-time behavior of an otherwise isolated system of 
many coupled spins. This question is a part of a more general problem of the evolution of a complex system in an external field,
especially in an intense external field. We hope that the methods here developed may be applied in these cases 
with the suitable modifications.
%

%
\appendix
\section{ EVOLUTION OF A SYSTEM IN AN ALTERNATING EXTERNAL FIELD}
\label{ef}
In section (\ref{nsok})  we wrote the kinetic and evolution equations in the approach of the
nonequilibrium statistical operator. In this Appendix we show briefly the derivation of the same equations in the
presence of   alternating external field. This problem is essential for the nuclear and electron spin resonance. Both
nuclear and electron spins have associated magnetic dipole moments which can absorb radiation, usually at
radio or microwave frequencies.\\
We  consider the many-particle system with the Hamiltonian
\begin{equation}\label{b1}
 H = H_{1} + H_{2} + V + H_{f}(t),
\end{equation}
where
\begin{equation}\label{b2}
 H_{1} = \sum_{\alpha} E_{\alpha}a^{\dagger}_{\alpha}a_{\alpha}
\end{equation}
is the single-particle second-quantized Hamiltonian of the quasiparticles with energies $ E_{\alpha}$. This term corresponds to the
kinetic energy of noninteracting particles $$H_{1} = \sum^{N}_{i=1} \frac{P_{i}^{2}}{2m} = 
\sum^{N}_{i=1}H(i), \quad H(i) = - \frac{\hbar^{2}}{2m} \nabla_{i}^{2}$$  
The index $\alpha \equiv ( \vec{k}, s)$
denotes the momentum and spin  $$ \varphi_{\alpha}(x) = \varphi_{\vec{k}}(\vec{r})\Delta(s-\sigma) 
= \exp (i \vec{k}\vec{r})\Delta(s-\sigma)/\sqrt{v} $$
$$E_{\alpha} = \langle \alpha| H_{1}| \alpha  \rangle$$ 
$$\langle k|H(1)|k' \rangle = \frac{1}{v} \int d^{3}r \exp (i \vec{k}\vec{r}) 
(- \frac{\hbar^{2}}{2m} \nabla^{2}) \exp (i \vec{k'}\vec{r}) = \frac{\hbar^{2}k^{2}}{2m}\Delta(k - k')$$
\\
and
\begin{equation}\label{b3}  
 \quad V = \sum_{\alpha,\beta}\Phi_{\alpha \beta}a^{\dagger}_{\alpha}a_{ \beta}, \quad \Phi_{\alpha \beta} = 
 \Phi^{\dagger}_{\beta \alpha }.
\end{equation}
Operator $V$ is the
operator of the interaction between the small subsystem and the thermal bath, and $H_{2}$ is the
Hamiltonian of the thermal bath  which we do not write explicitly. The quantities $\Phi_{\alpha \beta}$ are
the operators acting on the thermal bath variables with the properties $(\Phi_{\alpha \beta})^{\dagger} = 
\Phi^{*}_{ \beta \alpha}; \quad  \Phi^{*}_{ \beta \alpha} = \Phi_{\alpha \beta}$.\\
The interaction of the system with the external time dependent alternating field is described by the
operator
\begin{equation}\label{b4}  
H_{f}(t) = \sum_{\alpha,\beta}h_{\alpha \beta}(t)a^{\dagger}_{\alpha}a_{ \beta}, 
\end{equation}
For purposes of calculation, it is convenient to rewrite Hamiltonian $H_{f}(t)$ in a somewhat different form
\begin{equation}\label{b5}  
H_{f}(t) = \frac{1}{v} \sum_{\alpha,\beta}T (\alpha, \beta,t)a^{\dagger}_{\alpha}a_{ \beta}, 
\end{equation}
where
$$ h_{\alpha \beta}(t) = \frac{1}{v} T (\alpha, \beta,t)$$
and
$$ T = \sum_{i=1}^{N}T (\vec{r_{i}},t); \quad T (\vec{p}) = \int d^{3}r \exp (i \vec{p}\vec{r}) T (\vec{r},t)$$
$$\langle k|T (\vec{r},t)|k' \rangle = \frac{1}{v} \int d^{3}r \exp (i (\vec{k} - \vec{k'})\vec{r})T (\vec{r},t)
 = \frac{1}{v}
T (\vec{k} - \vec{k'},t) $$
\\
We are interested in the kinetic stage of the
nonequilibrium process in the system weakly coupled to the thermal bath. Therefore, we assume that the state of this
system is determined completely by the set of averages
$\langle P_{\alpha \beta}\rangle = \langle a^{\dagger}_{\alpha}a_{ \beta}\rangle$ and the state of the thermal bath 
by $\langle H_{2}\rangle$,
 where $ \langle \ldots \rangle$ denotes the statistical average with the nonequilibrium statistical operator,
 which will be defined below.\\
Following Pokrowsky's calculations we can write down the nonequilibrium statistical operator in the 
following form:
\begin{eqnarray}\label{b6}
\rho (t) = Q^{-1} \exp (- L(t)),
\end{eqnarray}
where
\begin{align}\label{b7}
L(t) = \sum_{\alpha \beta } P_{\alpha \beta } F_{\alpha \beta }(t) + \beta \mathcal H_{2}  \\
- \int^{0}_{-\infty} dt_{1} e^{\varepsilon t_{1}}
\left( \sum_{\alpha \beta } \dot{P}_{\alpha \beta }(t_{1}) F_{\alpha \beta}(t + t_{1}) +  
\sum_{\alpha \beta} P_{\alpha \beta }(t_{1}) \frac{\partial F_{\alpha \beta }(t + t_{1})}{\partial t_{1}}
+ \beta J_{2}\right)  \nonumber
\end{align}
The notation $\mathcal H_{2}$ denotes $\mathcal H_{2} = H_{2} - \mu_{2}N_{2}$ where $\mu_{2}$ is the
chemical potential of the medium (thermal bath) and $J_{2} = \dot{\mathcal H _{2}}(t_{1})$.
In this equation, the time dependence of the operators in the right-hand side differs from the time dependence
in Eq.(\ref{eq44a}). Consider this question in detail. The Heisenberg representation
\begin{equation}\label{b8}
  H(t) = U^{\dag}(t)H U(t); \quad U(t)= \exp (\frac{-iHt}{\hbar})
\end{equation}
in the presence of the external field $ H = H_{0} + H^{ext}$   takes the form
\begin{equation}\label{b9}
  A(t_{1};t) = U^{\dag}(t + t_{1};t)A U(t + t_{1};t)
\end{equation}
where
\begin{eqnarray}\label{b10}
U(t + t_{1};t) = T \exp (- \frac{i}{\hbar} \int_{t}^{t +t_{1}}H^{ext}(\tau) d\tau) 
\end{eqnarray}
We now generalize the evolution equations to the case in which the external field is present. 
We have
\begin{eqnarray}\label{b11}
\dot P_{\alpha \beta} = \frac{1}{i\hbar} [P_{\alpha \beta}, H ] = \nonumber \\ \frac{1}{i\hbar} (E_{\beta} - E_{\alpha})P_{\alpha \beta}
+ \frac{1}{i\hbar} \sum_{\nu} \left( P_{\alpha \nu}h_{\beta \nu}(t)  - h_{\nu \alpha}(t)P_{\nu \beta} \right) +
 \frac{1}{i\hbar}  [P_{\alpha \beta}, V ]
\end{eqnarray}
Then we can write down the balance equation
\begin{equation}\label{b12}
  J_{1}  + J_{2} = I_{f}, 
\end{equation}
where
\begin{equation}\label{b13}
  J_{1} = \dot H_{1} = \frac{1}{i\hbar} [H_{1}, H ] =  \frac{1}{i\hbar} ( [H_{1}, V ] + [H_{1}, H^{ext} ])
\end{equation}
and
\begin{equation}\label{b14}
I_{f} = J_{1}  + J_{2} =   \frac{1}{i\hbar} \sum_{\alpha \beta}\{ (E_{\alpha} - E_{\beta})h_{\alpha \beta}(t) +
h_{\alpha \beta}(t) [P_{\alpha \beta}, V ] \}
\end{equation}
The last term describes the work of the external field.\\  The parameters $F_{\alpha\beta }(t)$ are determined from the condition 
$\langle P_{\alpha \beta }\rangle = \langle P_{\alpha \beta }\rangle_{q}$.
The quasi-equilibrium statistical operator   $\rho_{q}$ has the form
\begin{equation}\label{b15}
 \rho_{q}  = \rho _{1}\rho_{2},
\end{equation}
where
\begin{eqnarray}\label{b16}
\rho _{1} = Q^{-1}_{1}\exp \left(- \sum_{\alpha \beta } P_{\alpha \beta } F_{\alpha \beta }(t)\right);  \nonumber \\
Q_{1} = Tr \exp \left( - \sum_{\alpha \beta } P_{\alpha \beta } F_{\alpha \beta }(t) \right)\\
\rho _{2} = Q^{-1}_{2} \exp \left( - \beta (H_{2} - \mu_{2}N_{2}) \right); \nonumber \\  
Q_{2} = Tr \exp \left(- \beta (H_{2}- \mu_{2}N_{2}) \right) \label{b17}
\end{eqnarray}
Thus, we can write
\begin{align}\label{b18}
\frac{d \langle P_{\alpha \beta }\rangle }{d t} =  \frac{1}{i \hbar}\langle [  P_{\alpha \beta  } , H ]\rangle = \\ \nonumber
 \frac{1}{i \hbar}(E_{\beta} - E_{\alpha})\langle P_{\alpha \beta  }\rangle   
+ \frac{1}{i\hbar} \sum_{\nu} \left(\langle P_{\alpha \nu} \rangle h_{\beta \nu}(t)  - 
h_{\nu \alpha}(t)\langle P_{\nu \beta} \rangle \right) +
 \frac{1}{i\hbar} \langle [P_{\alpha \beta}, V ] \rangle
\end{align}
To calculate explicitly the r.h.s. of  Eq.(\ref{b18}) we use the formula  
\begin{align}
  \exp (-A - B) = \exp (-A) - \int_{0}^{1}\exp (-A) (\exp (-A\tau) B \exp (A\tau) d\tau)\\
 \rho \simeq \{ 1 -  \int_{0}^{1} (\exp (-A\tau) B \exp (A\tau) -  
 \langle \exp (-A\tau) B \exp (A\tau)\rangle_{A} ) d\tau) \}\rho(A),
\end{align}
where
\begin{eqnarray*}
\rho(A) = \frac{\exp (-A)}{Tr \exp (-A)}\\
A = \sum_{\alpha \beta } P_{\alpha \beta } F_{\alpha \beta }(t) + \beta (H_{2}- \mu_{2}N_{2});\\
B = - \int_{-\infty}^{0} dt_{1} e^{\varepsilon t_{1}} \frac{1}{i\hbar} \sum_{\mu\nu}[P_{\mu \nu}(t_{1},t ), V(t_{1}) ]X_{\mu \nu}(t + t_{1})
\end{eqnarray*}
Making the same expansion procedure as described in section \ref{nsok} we find
\begin{align}\label{b19}
\frac{d \langle P_{\alpha \beta }\rangle }{d t} =  
 \frac{1}{i \hbar}(E_{\beta} - E_{\alpha})\langle P_{\alpha \beta  }\rangle +  \\ \nonumber
+ \frac{1}{i\hbar} \sum_{\nu} \left(\langle P_{\alpha \nu} \rangle h_{\beta \nu}(t)  - 
h_{\nu \alpha}(t)\langle P_{\nu \beta} \rangle \right) +
\frac{\beta}{i\hbar} \int^{1}_{0} d\lambda \langle [P_{\alpha \beta}, V ] e^{-\lambda A}V e^{\lambda A}\rangle_{q} + \\ \nonumber
\frac{1}{(i \hbar)^{2}} \int^{1}_{0} d\lambda \int_{-\infty}^{0} dt_{1} e^{\varepsilon t_{1}}
\sum_{\alpha'\beta'}\langle [P_{\alpha' \beta'}, V ] e^{-\lambda A}[P_{\alpha' \beta'}(t_{1},t ), V(t_{1}) ] e^{\lambda A}\rangle_{q}
X_{\alpha' \beta'}(t + t_{1}),
\end{align}
where
\begin{equation}\label{b20}
X_{\alpha' \beta'}(t ) =  F_{\alpha' \beta'}(t ) - \beta [ \delta_{\alpha' \beta'} \widetilde{E}_{\alpha'} 
+ h_{\alpha' \beta'}(t)]
\end{equation}
It can be rewritten as
\begin{align}\label{b21}
\frac{d \langle P_{\alpha \beta }\rangle }{d t} =  
 \frac{1}{i \hbar}(E_{\beta} - E_{\alpha})\langle P_{\alpha \beta  }\rangle +  \\ \nonumber
 + \frac{1}{i\hbar} \sum_{\nu} \left(\langle P_{\alpha \nu} \rangle h_{\beta \nu}(t)  - 
h_{\nu \alpha}(t)\langle P_{\nu \beta} \rangle \right) + \sum_{\alpha'\beta'}
K^{h}_{\alpha \beta,\alpha'\beta'} \langle P_{\alpha'\beta'}\rangle 
\end{align}
where the generalized    {\it relaxation matrix}   is given by
\begin{align}\label{b22} 
K^{h}_{\alpha \beta,\alpha'\beta'} = 
\frac{\beta}{i\hbar} \int^{1}_{0} d\lambda \langle [P_{\alpha \beta}, V ] e^{-\lambda A}V e^{\lambda A}\rangle_{q} + \\ \nonumber
\frac{1}{(i \hbar)^{2}} \int^{1}_{0} d\lambda \int_{-\infty}^{0} dt_{1} e^{\varepsilon t_{1}}
\langle [P_{\alpha' \beta'}, V ] e^{-\lambda A}[P_{\alpha' \beta'}(t_{1},t ), V(t_{1}) ] e^{\lambda A}\rangle_{q}
X_{\alpha' \beta'}(t + t_{1})
\end{align}
Equation (\ref{b21}) gives the generalization of the  rate equation (\ref{eq76}) of the Redfield-type for the case of the external
alternating field. A more detailed investigation of this equation and the problem of the evolution of a system in 
an external field will be carried out separately.
%
%
%
\section{ CORRELATION FUNCTIONS AND GAUSSIAN APPROXIMATION}
\label{ga}
The eigenvalues of the Hamiltonian  (\ref{eq107}) correspond to well-defined values of $\sum_{i}I^{z}_{i} = I^{z} = m$. Their energy
is the sum of a Zeeman energy $m \hbar\omega_{n}$ and a spin-spin energy. A stochastic-theoretical treatment of the
spin relaxation phenomena is a useful complementary approach to the consideration of spin evolution~\cite{kub,kub1}. 
By a stochastic theory it is termed ordinary that kind of
theoretical treatment of the problem in which one assumes the random nature of the forces acting on a system.
The phenomenon of spin relaxation can be properly
interpreted as some stochastic process of spin motion. 
This stochastic process is determined by the equation of motion of the spin variable. 
It was formulated~\cite{abr,kub} plausibly that a Gaussian random process may be well applied for the evolution
of the magnetization in the presence of a static external field 
\begin{equation}\label{a1}
  \frac{d}{dt } \vec{\mu}  = \gamma \vec{\mu}\times ( \vec{h_{0}} + \vec{h}),
\end{equation}
where $\gamma$ denotes the gyromagnetic ratio, $\vec{h_{0}}$ a static external field, and $\vec{h}$ the fluctuating
internal field due to the magnetic moments in the surrounding medium. The effect of the fluctuating internal field
$\vec{h}$ is to cause nuclear spin transitions governed by the selection rule $ \Delta m = \pm 1$. If the Zeeman 
splitting is small, i.e., $\hbar \omega_{n} \ll kT$, then the transition probability for a $ \Delta m = \pm 1$ transition
will be proportional to the Fourier transform of  correlation functions of the form $(h^{+}(t) h^{-}(t')), 
(h^{-}(t) h^{+}(t')), (h^{z}(t) h^{z}(t'))$. If we  assume the process of $\vec{h}(t)$ to be a Gaussian random
process, the problem becomes  more easy tractable.
From this viewpoint  it is reasonable to assume that the equation of spin motion involves the local fluctuating magnetic field
whose process is assumed to be a Gaussian random process~\cite{kub}. \\
The Gaussian or normal probability distribution  law is the limit of the binomial distribution
$$ P(m) = C^{m}_{n}p^{m}(1 - p)^{n}$$
in the limit of large $n$ and $pn$ ($n \rightarrow \infty$).
Here $n$ is the repetition of an experiment, $p$ is the probability of success, and $C^{m}_{n} = n!/m!(n - m)! $.
The normal probability distribution  has the form
\begin{equation}\label{a2}
 P(m) = \frac{1}{\sqrt{2\pi} \sigma} \exp \left( - \frac{1}{2} \frac{\xi^{2}}{\sigma^{2}}\right),
\end{equation}
where $\sigma = \sqrt{np(1 - p)}$ is a measure of the width of the distribution. It is clear that the Gaussian
distribution results when an experiment with a finite probability of success is repeated a very large number of times.
The Gaussian random process is a random process (with discrete or continuous time)  which has the normal (Gaussian)
probability distribution law for any group of values of the process. The Gaussian random process is determined completely
by its average value and correlation function. Thus, the description of the class of Gaussian processes is reduced to the
determination of the possible form of the corresponding correlation functions.\\
Consider now an isotropic distribution of nuclei and rewrite the production of the current operator in Eq.(\ref{eq106a})
in explicit form
\begin{eqnarray}\label{a3}
J( \vec{r} ) J( \vec{r_{1}},t_{1} ) = \frac{\omega_{n}^{2}}{4}\sum_{k \neq l}\sum_{m \neq n}
A_{kl}A_{mn}r_{kl}r_{mn} \\ \nonumber
\int d\vec{r}\delta(\vec{r} - \vec{r_{k}}) \delta(\vec{r_{1}} - \vec{r_{m}}) [Tr (I^{z})^{2} ]^{-1}
Tr I^{+}_{k}I^{-}_{l}I^{+}_{m}(t_{1})I^{-}_{n}(t_{1})
\end{eqnarray}
To proceed further, the form of the correlation function of nuclear spins in the above expression must be
determined. In the theory of NMR the reasonable assumption is that  this correlation function can be represented
in an intuitively understandable way as~\cite{bu65,kub}
\begin{eqnarray}\label{a4}
Tr I^{+}_{k}I^{-}_{l}I^{+}_{m}(t)I^{-}_{n}(t) \propto Tr (I^{+}I^{-})^{2}f(t) \delta_{kn} \delta_{lm}
\exp [ \frac{it}{\hbar}(\Omega_{l} - \Omega_{k} )] = \nonumber \\
\frac{1}{4}f(t) \delta_{kn} \delta_{lm} \exp [ \frac{it}{\hbar}(\Omega_{l} - \Omega_{k} )]
\end{eqnarray}
Then the diffusion coefficient $D(\vec{r} )$  (\ref{eq106a}) takes the form
\begin{eqnarray}\label{a5}
D(\vec{r} ) = \frac{1}{8\hbar^{2} N(r)} \sum_{k \neq l} A^{2}_{kl} r^{2}_{kl}\delta(\vec{r} - \vec{r_{k}})
\int^{\infty}_{- \infty} e^{\varepsilon t}dt f(t) \exp [ \frac{it}{\hbar}(\Omega_{k} - \Omega_{l} )] \nonumber \\
= \frac{1}{8\hbar^{2}} \sum_{l} A^{2}_{rl} r^{2}_{rl}
\int^{\infty}_{- \infty} dt f(t) \exp [ \frac{it}{\hbar}(\Omega_{r} - \Omega_{l} )].
\end{eqnarray}
The method of moments gives that $f(t)$ is close to the normal probability distribution 
\begin{equation}\label{a6}
 f(t) = A \exp ( - \frac{t^{2}\omega_{d}^{2}}{2}); \quad \hbar^{2}\omega_{d}^{2} = \frac{Tr H^{2}_{dip}}{Tr (I^{z})^{2}}
\end{equation}
The constant $A$ can be determined from the condition
\begin{equation}\label{a7}
 \int^{\infty}_{- \infty} dt f(t) = 1 = A \sqrt{\frac{2\pi}{\omega_{d}^{2}}}; \quad A = \sqrt{\frac{\omega_{d}^{2}}{2\pi}}
\end{equation}
Thus, we obtain
\begin{equation}\label{a8}
f(t) = \frac{\omega_{d}}{\sqrt{2\pi}} \exp ( - \frac{t^{2}\omega_{d}^{2}}{2})
\end{equation}
For the diffusion coefficient (\ref{a5}) we find
\begin{equation}\label{a9}
 D (\vec{r} ) \approx \frac{\omega_{d}}{\hbar^{2}\sqrt{\pi}} \sum_{l} A^{2}_{rl} r^{2}_{rl} 
 \exp [- (\Omega_{r} - \Omega_{l} )^{2}/4\ (\omega_{d})^{2}]. 
\end{equation}
In the case when $r$ is close to $l$  the frequency difference $(\Omega_{r} - \Omega_{l} ) \gg \omega_{d}$ and 
$ D (\vec{r} ) \rightarrow 0$. In the opposite case, when $(\Omega_{r} - \Omega_{l} ) \ll \omega_{d}$ the diffusion
coefficient is nearly constant $ D (\vec{r} ) \sim D$. Thus, we traced back to the notion of the diffusion barrier
$\delta$ (\ref{eq122}).
Consider two neighboring nuclei  along the radius
from the impurity. The distance between them is equal to the lattice constant $a$. For this case the frequency shift
is equal to $$(\Omega_{\delta} - \Omega_{\delta + a}) \approx \omega_{d},$$ 
where $\omega_{d} \approx 6 \gamma_{n}^{2} \hbar a^{-3}$. \\ Consider this constraint more carefully. We have
\begin{align}\label{a10}
\gamma_{n}\gamma_{M}\hbar \langle S^{z} \rangle \{ \frac{1 + B  \cos (\delta k_{F} + \phi_{B} )}{\delta^3} -
\frac{1 + B  \cos ((\delta + a ) k_{F} + \phi_{B} )}{(\delta + a)^3}\} = \nonumber \\
\gamma_{n}\gamma_{M}\hbar \langle S^{z} \rangle \{\frac{1 + B  \cos (\delta k_{F} + \phi_{B} )}{\delta^3} -
\frac{1 - B  \sin (\delta k_{F} + \phi_{B} )a k_{F}}{\delta^3} + \nonumber \\ +
 \frac{B  \cos (\delta k_{F} + \phi_{B} )}{\delta^3} + 3  \frac{1 + B  \cos (\delta k_{F} + \phi_{B} )}{\delta^4}a \}
= \nonumber \\
\frac{a}{\delta^{3}} \gamma_{n}\gamma_{M}\hbar \langle S^{z} \rangle \{ 
3 \frac{1 + B  \cos (2\delta k_{F} + \phi_{B} )}{\delta} - \nonumber \\
B \sin (\delta k_{F} + \phi_{B} )k_{F} \} =  
6 \gamma_{n}^{2} \hbar a^{-3}
\end{align}
For the rough estimation we omit the $\cos$ and $\sin$ contributions. Then we obtain
\begin{equation}
6\gamma_{n}^{2}\hbar a^{-3} = \gamma_{n}\gamma_{M}\hbar  \langle S^{z} \rangle \frac{a}{\delta^4}; \quad
\delta = a \sqrt[4]{[ \frac{\gamma_{M}}{\gamma_{n}}  \langle S^{z} \rangle ]}
\end{equation}
%
%
%
%
%
%

%

\begin{thebibliography}{9}
%
%
%
\bibitem{cas37}
H. B. G. Casimir   and F. K. Du Pre, {\it Physica}  {\bf 5}, 507 (1937).
%
%
%
\bibitem{cas39}
H. B. G. Casimir, {\it Physica}  {\bf 6}, 156 (1939).
%
%
\bibitem{bloc}
F. Bloch, {\it  Phys.  Rev.}   {\bf 70}, 460 (1946).
%
%
%
\bibitem{gor} C. J. Gorter,
{\em Paramagnetic Relaxation  }, Elsevier   (Amsterdam, New York,  1947).
%
%
%
\bibitem{bl49}
N. Bloembergen, {\it Physica}  {\bf 15}, 386
(1949).
%
%
\bibitem{kt}
R. Kubo and K. Tomita, {\it J. Phys.  Soc. Jpn } {\bf 9}, 888 (1954).
%
%
\bibitem{red56}
A. G. Redfield,  {\it Phys.  Rev.}   {\bf 101}, 67 (1956).
%
%
\bibitem{red57}
A. G. Redfield,  {\it IBM J. Res. Develop.}   {\bf 1}, 19 (1957).
%
%
%
\bibitem{red59}
A. G. Redfield, {\it  Phys.  Rev}.   {\bf 116}, 315 (1959).
%
%
\bibitem{hsl}
L. C. Hebel and C. P. Slichter,  {\it Phys.  Rev}.   {\bf 113}, 1504 (1959).
%
%
%
%
\bibitem{ca60}
W. J. Caspers, {\it Physica}  {\bf 26}, 778, 798 (1960).
%
%
\bibitem{sher}
A. Sher and H. Primakoff,  {\it Phys.  Rev}.   {\bf 119}, 178 (1960).
%
%
\bibitem{sherp}
A. Sher and H. Primakoff,  {\it Phys.  Rev}.   {\bf 130}, 1267 (1963).
%
%
%
\bibitem{blo} N. Bloembergen,
{\em Nuclear Magnetic Relaxation  }, W. A. Benjamin   (New York,  1961).
%
%
\bibitem{kla}
J. R. Klauder and P. W. Anderson,  {\it Phys.  Rev}.   {\bf 125}, 912 (1962).
%
%
\bibitem{ko}
J. Korringa,   {\it Phys.  Rev}.   {\bf 133}, A1228 (1964).
%
%
\bibitem{ko64}
J. Korringa, J. L. Motchane, P. Papon and A. Yashimori,  {\it Phys.  Rev}.   {\bf 133}, A1230 (1964).
%
%
\bibitem{sten}
S. Stenholm and D. ter Haar, {\it Physica}  {\bf 32}, 1361 (1966).
%
%
%
\bibitem{cas64} W. J. Caspers,
{\em Theory of Spin Relaxation  }, Interscience Publ.   (New York,  1964).
%
%
\bibitem{hanh}
R. L. Strombothe and E. L. Hanh,  {\it Phys.  Rev}.   {\bf 133}, A1616 (1964).
%
%
\bibitem{a63}
P. N. Argyres, in: {\it Proc. Eindhoven Conf. on Magnetic and Electric Resonance and Relaxation}, edited by J. Smidt,
North-Holland ( Amsterdam , 1963), p.555.
%
%
\bibitem{ak}
P. N. Argyres and P. L. Kelley,  {\it Phys.  Rev}.   {\bf 134}, A98 (1964).
%
%
\bibitem{bu65}
L. L. Buishvili and D. N. Zubarev,  {\it Solid State Physics},
{\bf 7}, 722 (1965).
%
%
%
\bibitem{sa66}
G. Sauermann, {\it Physica } {\bf 32}, 2017
(1966).
%
%
%
\bibitem{red66}
A. G. Redfield, in:  {\it Adv. Magn.  Resonance}, edited by J. S. Waugh   (New York, 1966),  Vol. {\bf 1}, p.1.
%
%
%
%
\bibitem{lo67}
L. J. Lowe and S. Gade,  {\it Phys.  Rev}.   {\bf 156}, 817 (1967).
%
%
\bibitem{rob}
B. Robertson, {\it  Phys.  Rev}.   {\bf 153}, 391 (1967).
%
%
%
\bibitem{jen68}
J. Jeener, in {\it Adv. Magn.  Resonance}, edited by J. S. Waugh   (New York, 1968),  Vol. {\bf 3}, p.206.
%
%
\bibitem{mora}
P. R. Moran,  {\it J. Phys. Chem. Solids}, {\bf 30}, 297 (1969).
%
%
\bibitem{hm}
H. Meyer, S. M. Myers and J. P. Remeika,  {\it J. Phys. Chem. Solids}, {\bf 30}, 2687 (1969).
%
%
\bibitem{hm1}
H. Meyer, S. M. Myers and J. P. Remeika, {\it  J. Phys. Chem. Solids}, {\bf 32}, 867 (1971).
%
%
%
%
\bibitem{gold} M. Goldman,
{\em Spin Temperature and Nuclear Magnetic Resonance in Solids }, Oxford University Press   (Oxford, New York,  1970).
%
%
%
%
\bibitem{abr} A. Abragam,
{\em Principles of Nuclear Magnetism }, Oxford University Press   (Oxford,   1961).
%
%
%
\bibitem{wolf} D. Wolf,
{\em Spin-temperature and Nuclear-spin  Relaxation in Matter }, Oxford University Press   (Oxford, New York,  1979).
%
%
%
\bibitem{sl} C. P. Slichter,
{\em Principles of Magnetic Resonance}, Springer   ( Berlin , 1980).
%
%
%
\bibitem{lou}
G. J. Ehnholm, J. P. Ekstr\"{o}m, J. F. Jacquinot, M. J. Loponen, O. V. Lounasmaa and J. K. Soini,  {\it J. Low Temp. Phys. } {\bf 39}, 417
(1980).
%
%
%
\bibitem{ab} A. Abragam and M. Goldman,
{\em  Nuclear Magnetism: Order and Disorder }, Oxford University Press   (Oxford, New York,  1982).
%
%
%
\bibitem{bo04}
G. S. Boutis, D. Greenbaum, H. Cho, D. G. Gory and C. Ramanathan, {\it Phys.  Rev. Lett.} {\bf 92}, 137201
(2004).
%
%
\bibitem{pin}
W. K. Rhim, A. Pines and J. S. Waugh,  {\it Phys.  Rev. Lett.} {\bf 25}, 218
(1970).
%
%
\bibitem{buis65}
L. L. Buishvili,  {\it Zh.Exp.Th. Phys.},
{\bf 49}, 1868 (1965).
%
%
\bibitem{bui65}
L. L. Buishvili,  {\it Solid State Physics},
{\bf 7}, 1871 (1965).
%
%
\bibitem{bbz}
N. S. Bendiashvili, L. L. Buishvili and M. D. Zviadadze,  {\it Solid State Physics},
{\bf 8}, 2919 (1966).
%
%
\bibitem{buzv}
L. L. Buishvili and M. D. Zviadadze,  {\it Phys.Lett.} A
{\bf 25}, 86 (1967).
%
%
\bibitem{buzv68}
L. L. Buishvili and M. D. Zviadadze,  {\it Solid State Physics}  
{\bf 10}, 2553 (1968).
%
%
\bibitem{bowa} P. Borckmans and D. Walgraef,
in: {\it Proc. XV Colloque Ampere,} ,
North-Holland ( Amsterdam , 1969), p.418.
%
%
\bibitem{bug69}
L. L. Buishvili and N. P. Giorgadze,  {\it Doklady Academii Nauk SSSR}  
{\bf 189}, 508 (1969).
%
%
\bibitem{bzkh69}
 L. L. Buishvili, M. D. Zviadadze and G. R. Khutsishvili,  {\it Zh.Exp.Th. Phys.},
{\bf 56}, 290 (1969).
%
%
\bibitem{bug70}
L. L. Buishvili and N. P. Giorgadze,  {\it Solid State Physics}  
{\bf 12}, 1817 (1970).
%
%
\bibitem{bbz70}
N. S. Bendiashvili, L. L. Buishvili and M. D. Zviadadze,  {\it Zh.Exp.Th. Phys.},
{\bf 58}, 597 (1970).
%
%
\bibitem{bz70}
L. L. Buishvili and N. P. Giorgadze,  {\it Doklady Academii Nauk SSSR}  
{\bf 191}, 58 (1970).
%
%
\bibitem{faz}
N. G. Fazleev,  {\it Sov. J. Low Temp. Phys. } {\bf 5}, 181   (1979).
%
%
\bibitem{faz1}
N. G. Fazleev,  {\it Sov. J. Low Temp. Phys. } {\bf 6}, 693   (1980).
%
%
\bibitem{tay}
D. Tayurskii, {\it phys.stat.sol.}(b)  {\bf 162}, 545 (1989).
%
%
\bibitem{nigm}
R. Nigmatullin and D. Tayurskii, {\it Physica A}  {\bf 175}, 275 (1991).
%
%
\bibitem{rom}
V.Romero-Rochin, A. Orsky and I. Oppenheim, {\it Physica A}  {\bf 156}, 244 (1989).
%
%
%
\bibitem{grand87} W. T. Grandy,
{\em Foundations of Statistical Mechanics: Equilibrium Theory  }, vol.1,  D. Reidel Publ.   (Doddrecht, Holland,  1987).
%
%
%
\bibitem{grand88} W. T. Grandy,
{\em Foundations of Statistical Mechanics:  Nonequilibrium Phenomena  }, vol.2, D. Reidel Publ.   (Doddrecht, Holland,  1988).
%
%
\bibitem{kubo92} M. Toda, R. Kubo and N. Saito,
{\em  Statistical Physics: Equilibrium Statistical Mechanics  }, vol.1, Springer Publ.   (Berlin,  1992).
%
%
%
\bibitem{kubo91} R. Kubo, M. Toda and N. Hashitsume,
{\em  Statistical Physics: Nonequilibrium Statistical Mechanics  }, vol.2, Springer Publ.   (Berlin,  1991).
%
%
%
\bibitem{zub74} D. N. Zubarev,
{\em Nonequilibrium Statistical Thermodynamics  }, Consultant Bureau   (New York,  1974).
%
%
%
%
\bibitem{macl89} J. A. McLennan
{\em Introduction to Nonequilibrium Statistical Mechanics }, Prentice Hall  (New Jersey,  1989).
%
%
%
\bibitem{rz} R. Zwanzig,
{\em Nonequilibrium Statistical Mechanics  }, Oxford University Press   (Oxford, New York,  2001).
%
%
%
\bibitem{leb}
J. L. Lebowitz, {\it Physica  A } {\bf 263}, 516 (1999).
%
%
\bibitem{kuz05}
A. L. Kuzemsky, {\it Intern. J. Modern Phys. B } {\bf 19}, 1029 (2005). (cond-mat/0502194)
%
%
%
%
%
%
%
\bibitem{bog} N. N. Bogoliubov,
%
in: {\em Studies in  Statistical Mechanics  }
 edited by J. de  Boer and G. E. Uhlenbeck,
North-Holland ( Amsterdam , 1962), Vol. {\bf 1}, p.1.
%
%
%
\bibitem{bog1} N. N. Bogoliubov,``On the stochastic processes in the dynamical systems'',
{\it Communications E17-10514}, JINR, Dubna (1977).
%
%
%
%
\bibitem{zbig}
Z. Onyszkiewicz, {\it Physica  A}    {\bf 143}, 287 (1987).
%
%
%
%
\bibitem{gell}
M. Gell-Mann and M. L. Goldberger,  {\it Phys.  Rev.}   {\bf 91}, 398 (1953).
%
%

\bibitem{qbog} N. N. Bogoliubov,``Quasi-averages in
the problems  of statistical mechanics'',
{\it Communications D-781}, JINR, Dubna (1961).
%
%
%
%
%
\bibitem{vh}
L. van Hove, {\it Physica }  {\bf 23}, 411 (1957).
%
%
%
\bibitem{zw}
R. Zwanzig, {\it Physica }  {\bf 30}, 1109 (1964).
%
%
%
\bibitem{fk}
A. Fulinski  and W. J. Kramarczyk, {\it  Physica }  {\bf 39}, 575 (1968).
%
%
\bibitem{pok}
L. A. Pokrowsky, {\it Doklady Academii Nauk SSSR } {\bf 183}, 806 (1968).
%
%
%
%
\bibitem{cd}
R. I. Cukier and J. M. Deutch, {\it  J. Chem. Phys.}  {\bf 50}, 36 (1969).
%
%
%
%
%
\bibitem{hub}
P. Hubbard, {\it  Rev. Mod. Phys}.  {\bf 33}, 249 (1961).
%
%
%
\bibitem{khu}
G. R. Khutsishvili, {\it  Usp. Fiz. Nauk. } {\bf 87}, 211 (1965); {\it Sov. Phys. Uspekhi}, {\bf 8}, 743 (1966).
%
%
\bibitem{khu1}
G. R. Khutsishvili,  {\it Comments on Solid State Physics } {\bf 5}, 23 (1972).
%
%
%
%
\bibitem{mf}
P. M. Morse and H. Feshbach,  {\em  Methods of Theoretical Physics }, McGraw-Hill Book Co.,   ( New York,  1953).
%
%
\bibitem{ar}
 G. Arfken,  {\em Mathematical Methods for Physicists }, Academic Press   ( New York,  1985).
%
%
\bibitem{sta}
 J. P. Stark,  {\em Solid State Diffusion }, J. Wiley and Sons   ( New York,  1976).
%
%
\bibitem{blum}
W. E. Blumberg,  {\it Phys.  Rev.}   {\bf 119}, 79 (1960).
%
%
%
\bibitem{rig}
D. A. Rigney and C. P. Flynn, {\it  Phil. Mag.}   {\bf 15}, 1213 (1967).
%
%
\bibitem{hir}
E. C. Hirschkoff, O. G. Symko and J. C. Wheatley,  {\it J. Low Temp. Phys. } {\bf 5}, 155
(1971).
%
%
%
\bibitem{pet}
J. J. Peters and C. P. Flynn,  {\it Phys.  Rev. } B {\bf 6}, 3343 (1972).
%
%
%
%
\bibitem{mor}
K. Yoshida and T. Moriya, {\it J. Phys.  Soc. Jpn } {\bf 11}, 33 (1956).
%
%
\bibitem{koh}
W. Kohn and S. H. Vosko,  {\it Phys.  Rev.}   {\bf 119}, 912 (1960).
%
%
%
\bibitem{walke}
M. B. Walker, {\it Phys.  Rev.}   {\bf 176}, 432 (1968).
%
%
\bibitem{gih}
B. Giovannini and A. J. Heeger,  {\it Helv. Phys. Acta }   {\bf 42}, 639 (1969).
%
%
%
%
\bibitem{js}
J. Solyom,  {\it Z. Phys.}     {\bf 238}, 195 (1970).
%
%
%
\bibitem{walk}
M. B. Walker,  {\it Phys.  Rev. } B {\bf 1}, 3690 (1970).
%
%
%
\bibitem{gi}
B. Giovannini, P. Pincus, G. Gladstone and A. J. Heeger,  {\it J. Physique }   {\bf 32}, C1-163 (1971).
%
%
%
%
\bibitem{zit}
J. Zitkova-Wilcox,  {\it Solid State Commun.}     {\bf 12}, 1109 (1973).
%
%
\bibitem{noz}
P. Nozieres,  {\it J. Low Temp. Phys. } {\bf 17}, 31 (1974).
%
%
%
\bibitem{br1}
N. Bloembergen and T. J. Rowland, {\it Acta Metallurgica}   {\bf 1}, 731 (1953).
%
%
\bibitem{br}
N. Bloembergen and T. J. Rowland, {\it Phys.  Rev.}   {\bf 97}, 1679 (1955).
%
%
\bibitem{row}
T. J. Rowland, {\it Phys.  Rev.}   {\bf 119}, 900 (1960).
%
%
%
\bibitem{stre}
R. L. Streever, {\it Phys.  Rev.}   {\bf 134}, A1612 (1964).
%
%
%
\bibitem{clo}
S. Clough and W. I. Goldburg,  {\it J. Chem. Phys.}     {\bf 45}, 4080 (1966).
%
%
%
\bibitem{ba} J. R. Asik, M. A. Ball and C. P. Slichter,
in: {\it Magnetic Resonance and Relaxation}, ({\it Proc. XIV Colloque Ampere, Ljubljana, 1966})  edited by R. Blinc,
North-Holland ( Amsterdam , 1967), p.448.
%
%
\bibitem{lum}
O. J. Lumpkin,  {\it Phys.  Rev.}   {\bf 164}, 324 (1967).
%
%
%
\bibitem{gos}
A. C. Gossard,  A. J. Heeger and J. H. Wernick,  {\it J. Appl. Phys.}     {\bf 38}, 1251 (1967).
%
%
\bibitem{wals}
R. E. Walstedt,  R. C. Sherwood and J. H. Wernick,  {\it J. Appl. Phys.}     {\bf 39}, 555 (1968).
%
%
%
\bibitem{wal}
R. E. Walstedt and A. Narath,   {\it Phys.  Rev. B}  {\bf 6}, 4118 (1972).
%
%
\bibitem{aa}
M. I. Aaalto, P. M. Berglund, H. K. Collan, G. J. Ehnholm, R. G. Gylling and O. V. Lounasmaa, {\it Physica }  {\bf 61}, 314 (1972).
%
%
\bibitem{lo}
D. C. Lo, D. V. Lang, J. B. Boyce and C. P. Slichter,   {\it Phys.  Rev. B}  {\bf 8}, 973 (1973).
%
%
\bibitem{all}
H. Alloul and P. Bernier,  {\it J. Phys. F: Metal Phys.} {\bf 4}, 870 (1974).
%
%
\bibitem{allo}
H. Alloul, J. Darville and P. Bernier,  {\it J. Phys. F: Metal Phys.} {\bf 4}, 2050 (1974).
%
%
\bibitem{bern}
P. Bernier and H. Alloul,  {\it J. Phys. F: Metal Phys.} {\bf 6}, 1193 (1976).
%
%
\bibitem{wawa}
R. E. Walstedt and L. R. Walker,   {\it Phys.  Rev. B}  {\bf 9}, 4857 (1974).
%
%
\bibitem{yaf}
Y. Yafet and V. Jaccarino,   {\it Phys.  Rev.}   {\bf 133}, A1630 (1964).
%
%
%
\bibitem{rk}
M. A. Ruderman and C. Kittel,   {\it Phys.  Rev.}   {\bf 96}, 99 (1954).
%
%
\bibitem{kuze05}
A. L. Kuzemsky, {\it Physica B}  {\bf 355}, 318 (2005). 
%
%
\bibitem{vv}
J. H. Van Vleck,   {\it Phys.  Rev.}   {\bf 74}, 1168 (1948).
%
%
\bibitem{coq} B. Coqblin
{\em The Electronic Structure of Rare-Earth Metals and Alloys: the
Magnetic Heavy Rare-Earths}, Academic Press  (New York, 1977).
%
%
\bibitem{mem}
F. Lado, J. D. Memory and G. W. Parker,   {\it Phys.  Rev. B}  {\bf 4}, 1406 (1971).
%
%
%
\bibitem{kor}
J.  Korringa, {\it Physica} {\bf 16}, 601 (1950). 
%
%
\bibitem{tmor}
T. Moriya, {\it J. Phys.  Soc. Jpn } {\bf 18}, 516 (1963).
%
%
\bibitem{kub}
R. Kubo, in: {\it Fluctuation, Relaxation, and Resonance in Magnetic Systems} ( Proc. Scottish Universities
Summer School, 1961 ), edited by D. ter Haar, Oliver and Boyd, 
 ( Edinburgh , 1962), p.23.
%
\bibitem{kub1}
R. Kubo, {\it Hyperfine Interactions}   {\bf 8}, 731 (1981).   
%
%
%
\end{thebibliography}
\end{document}